\documentclass[preprint,authoryear,12pt]{elsarticle}

\usepackage{epsfig}
\usepackage{amssymb}
\usepackage{graphicx}

\usepackage{color}
\newcommand{\cb}{}	
\newcommand{\cred}{}
\newcommand{\cc}{}	
\newcommand{\Vcorr}{}

\usepackage{amsmath} \usepackage{stmaryrd}
\usepackage{epic} \usepackage{eepic}
\usepackage{latexsym}

\usepackage{fancyhdr}

\renewcommand{\rho}{\varrho}
\renewcommand{\phi}{\varphi}
\newcommand{\nn}{\nonumber}

\newcommand{\beq}{\begin{equation}}
\newcommand{\eeq}{\end{equation}}
\newcommand{\bea}{\begin{eqnarray}}
\newcommand{\eea}{\end{eqnarray}}
\newcommand{\beas}{\begin{eqnarray*}}
\newcommand{\eeas}{\end{eqnarray*}}

\newcommand{\Images}{y}  

\newcommand{\mbs}[1]{\mathbf{#1}}

\def\bJ{{\mbs{J}}}

\def\bp{{\mbs{p}}}  \def\br{{\mbs{r}}}
  
  \def\bx{{\mbs{x}}}

\begin{document}

\begin{frontmatter}

\title{On the estimation of the curvatures and bending rigidity of membrane networks via a local maximum-entropy approach}

\date{today}

\author{F.~Fraternali}
\ead{fernando.fraternali@kcl.ac.uk}
\address{Department of Civil Engineering, University of Salerno,
84084 Fisciano(SA), Italy, and \\
Division of Engineering,
King's College London,
Strand, London  WC2R 2LS,
UK}

\author{C.~D.~Lorenz}
\ead{chris.lorenz@kcl.ac.uk}
\address{Department of Physics,
King's College London,
Strand, London  WC2R 2LS,
UK}

\author{G.~Marcelli* \footnote{*Corresponding author}}
\ead{gianluca.marcelli@kcl.ac.uk}
\address{Division of Engineering,
King's College London,
Strand, London  WC2R 2LS,
UK}

\begin{abstract}
We present a meshfree method for the curvature estimation of membrane networks based on the Local Maximum Entropy approach recently presented in \citep{Arrojo:2006}. A continuum regularization of the network is carried out by balancing the maximization of the information entropy corresponding to the nodal data, with the minimization of the total width of the shape functions. The accuracy and convergence properties of the given curvature prediction procedure are assessed through numerical applications to benchmark problems, {\cb{which include coarse grained molecular dynamics simulations of the fluctuations of red blood cell membranes \citep{Marcelli:2005, Hale:2009}.}} 
{\cred{We also provide an energetic discrete-to-continuum approach to  the prediction of the zero-temperature bending rigidity of membrane networks, which is based on the integration of the local curvature estimates.}}
The Local Maximum Entropy approach is easily applicable to the continuum regularization of fluctuating membranes,  and the prediction of membrane and bending elasticities of molecular dynamics models.
 \end{abstract}


\end{frontmatter}



\section{\label{intro}Introduction}
The estimation of the curvature tensor of membrane networks embedded in the 3D Euclidean space plays a key role in many relevant problems of  differential geometry, solid mechanics, biomechanics, biophysics and computer vision.
{\cb{Particularly important is the curvature estimation of fluctuating bio-membranes, which are often modeled as particle networks, via molecular dynamics (MD) and/or coarse grained molecular dynamics (CGMD) approaches.}}
The plasticity of cellular membranes is dependent on accurately selected mechanisms for sensing curvature and adopt different responses according to the particular membrane curvature.  
These mechanisms depend on the interplay between proteins and lipids and can be modulated by changes in lipid composition \citep{RisseladaetMarrink:2009}, membrane fusion \citep{Martensetal.:2008}, formation of raft-like domains, oligomerization of scaffolding proteins and/or insertion of wedge proteins into membranes.  
The dynamical changes in the membrane curvature can give rise to cell membrane remodelling \citep{McMahonetal.:2005} resulting in the formation of microenvironments that can facilitate the interaction of biomolecules in the cell.  On a larger scale, these dynamical changes play a key role in controlling cellular
growth, division and movement processes. Furthermore, as we already noticed, there has been a significant amount of modeling work focussed on characterizing the bending rigidity of ordered membranes  (refer, e.g., {\Vcorr to} \cite{Seung:1988}), vesicle membranes \citep{Gompper:1996, Du:2006} and the red blood cell membrane \citep{Marcelli:2005, Hale:2009, Suresh:2006}.   

The continuum regularization of a membrane network is naturally performed through meshfree approximation schemes, which are well suited for the discrete-to-continuum scale bridging.
Recently, a Local Maximum Entropy (LME) approach has been proposed to construct smooth meshfree approximants of given nodal data 
\citep{Arrojo:2006, Cyron:2009, Li:2010}. 
The LME approach is a convex, non-interpolant approximation method that
suitably balances the maximization of the information 
entropy corresponding to the given data \citep{Jaines:1957}, with the minimization of the 
total width of the shape functions
\citep{Rajan:1994}.
Some of the distinctive features of such an approach consist of the non-negativity and the partition of unity properties of the shape functions, which in particular can be thought of as the elements of a discrete probability distribution;
a weak interpolation (Kronecker-delta) property at the boundary;
{\cb{and first- or higher-order consistency conditions 
\citep{Arrojo:2006,Cyron:2009} }}.
As compared to popular, `explicit' approximation methods, such as approaches {\cb{utilizing}} B-Splines and Non-Uniform Rational B-Splines (NURBS), the LME requires more calculations and specifically the solution of a convex nonlinear optimization problem at each sampling point. 
Nevertheless, the LME guarantees high accuracy and smoothness of the continuous mapping \citep{Cyron:2009}, which are properties of fundamental importance when dealing with curvature estimation. 
{\cred{Due to its mixed, local-global character, the LME approximation scheme can be conveniently used to filter the inherent small scale roughness of the membrane, which is a distinctive feature of such an approach, as compared to popular computer graphic methods for estimating the curvatures of point-set surfaces,}} 
{\cb{(e.g. moving least-squares (MLS) methods). }}
{\cb{Another peculiar advantage of the LME regularization consists of its ability to handle unstructured node sets, which do not require any special pre-processing in such a scheme }}
{\cred{An extensive comparison}} {\cb{of the application of the}} {\cred{LME, MLS and B-Spline approaches to structural vibration problems is presented in \citep{Cyron:2009}.
}}

The present work deals with the formulation and the implementation of a curvature estimation method for membrane networks, which is based on the LME approach proposed in \citep{Arrojo:2006}. To the authors' knowledge, such an application of the LME approximation has not appeared in the literature yet.
In Section \ref{LME}, we provide {\cb{the}} explicit formulae for the second-order derivatives of the LME shape functions (not given in \citep{Arrojo:2006}), and a LME procedure for the estimation of the lines of curvature and the principal curvatures at the generic node of a membrane network. 
Next, we present in Section \ref{numerics} some numerical applications of the above procedure to membrane networks extracted from a sinusoidal surface and a spherical surface, establishing comparisons with exact solutions and assessing the convergence properties of the LME estimates. 
{\cred{We also provide, in the same 
section, some numerical results about the principal curvatures  of the red blood cell (RBC) model proposed in 
\citep{Marcelli:2005, Hale:2009}, and a discrete-to-continumm approach to the prediction of the bending rigidity at zero temperature of MD membrane networks. 
The method and results presented here represent an essential first step towards an extensive estimation of the elastic moduli of the RBC, which will be the specific subject of future studies. }}
Additionally, we plan to use the same approach to measure from trajectories of coarse-grain MD simulations
the curvature of cell membranes affected by asymetric lipid bilayers or by protein-lipids
interactions.
{\cb{Such future extensions of the present work are summarized in Section \ref{conclusions}, which also includes some final comments on the results presented in Section \ref{numerics}.}}

\bigskip

\section{\label{LME}LME regularization of membrane networks}

\subsection{Generalities on the LME approximation}

{\cb{First, we will present how to find the continuum regularization of a given discrete set ${ X}_N$ of $N$ nodes (or vertices) having Cartesian coordinates $\{ x_{a_1}, x_{a_2}, z_{a} \}$ ($a=1,...,N$ ) with respect to a given frame $\{O,x_1,x_2,z\equiv x_3\}$.}}
We wish to construct a continuum surface described by the Monge chart

\bea 
z \ = \ z_N(\bx) \ = \ \sum^{N}_{a=1}{ z_{a} p_{a}(\bx)},
\label{z_maxent}
\eea

{\cb{\noindent where $p_a$ are suitable \textit{shape functions} of the position vector $\bx=\{x_1,x_2\}$ in the $x_1,x_2$ plane. }}
When adopting the local maximum entropy approach proposed in \citep{Arrojo:2006}, we determine the functions $p_a$ by solving the following optimization problem:

\bigskip 
For a given $\bx$, {\cb{we minimize:}}
\bea 
f_\beta (\bx, \bp) \equiv \beta \sum^{N}_{a=1}{p_a \left|\bx-\bx_a\right|^2} + \sum^{N}_{a=1}{p_a \log p_a}
\label{1a}
\eea
subject to:
\bea 
p_a\geq0, \ a=1, ..., N
\label{1b}
\eea
\bea 
\sum^{N}_{a=1} p_a = 1
\label{1c}
\eea
\bea 
\sum^{N}_{a=1} p_a \bx_a = \bx
\label{1d}
\eea

\noindent where $\beta \in [0,+\infty)$ is a scalar parameter, and $\bp = \{p_1,..., p_N\}$. 
The constraints (\ref{1b}), (\ref{1c}) and (\ref{1d}) enforce the non-negativity of the shape functions, the partition of unity property, and the first-order consistency conditions, respectively. It is worth noting that equations (\ref{1c}) and (\ref{1d}) guarantee that affine functions are exactly reproduced by the LME  scheme  (\citep{Arrojo:2006,Cyron:2009}). 
On the other hand, equations (\ref{1b}) and (\ref{1c}) allow us to regard $\bp(\bx)$ as a discrete probability distribution, and the quantity $H_I(\bp) = - \sum^{N}_{a=1}{p_a \log p_a}$ as the corresponding information entropy (\citep{Jaines:1957}).
The quantity $W(\bp) = \sum^{N}_{a=1}{p_a \left|\bx-\bx_a\right|^2}$ instead represents the total width of the shape functions $p_a$ at the given $\bx$.
Depending on the value of $\beta$, the LME problem suitably balances the maximization of the information entropy corresponding to the given nodal data with the minimization of the total width of the shape functions $p_a$. A global maximum-entropy scheme (\citep{Jaines:1957}) is recovered by setting $\beta=0$ in (\ref{1a}), while a minimum-width  approximation scheme (\citep{Rajan:1994}) is obtained in the limit $\beta \rightarrow + \infty$.

{\cb{Now, we }} introduce the \textit{partition function}  $Z(\bx, \boldsymbol{\lambda})=\sum^{N}_{a=1}{Z_a(\bx, \boldsymbol{\lambda})}$, where $\boldsymbol{\lambda}=\{\lambda_1,\lambda_2\}$ denotes the vector of the Lagrange multipliers of the first-order consistency conditions (\ref{1d}), and it results in

\bea 
Z_a(\bx, \boldsymbol{\lambda}) = \exp\left[-\beta\left|\bx-\bx_a\right|^2+\boldsymbol{\lambda}\cdot (\bx-\bx_a)\right].
\label{7}
\eea

\noindent It can be shown (\citep{Arrojo:2006}) that, for any $\bx \in \mbox{conv} X$, the LME problem admits the unique solution shown below

\bea 
p^*_a(\bx) = \frac{Z_a(\bx, \boldsymbol{\lambda}^*)}{Z(\bx, \boldsymbol{\lambda})}
\label{6}
\eea

\noindent where 

\bea 
\boldsymbol{\lambda}^* =  \mbox{arg} \min_{\boldsymbol{\lambda} \in \Re^d} \left\{ F( \boldsymbol{\lambda}) = \log Z(\bx, \boldsymbol{\lambda}) \right\}
\label{11}
\eea

\noindent  It is useful to the employ the Newton-Raphson method to solve equation (\ref{11}) iteratively.  
Let $\boldsymbol{\lambda}^k$ denote the approximate solution to (\ref{11}) at the $k$th iteration. A Newton-Raphson update furnishes

\bea 
\boldsymbol{\lambda}^{k+1} = \boldsymbol{\lambda}^k \  \ 
-\left(\bJ^{-1}\right)^k \br^k,
\label{13}
\eea

\noindent  where $\br^k$ and $\bJ^K$ are the particularization of the gradient $\br$ and the Hessian  $\bJ$ of $F$ for $\boldsymbol{\lambda} = \boldsymbol{\lambda}^k$. Straightforward calculations (cf. \citep{Arrojo:2006}) 
give

\bea 
\br (\bx, \boldsymbol{\lambda}) = \nabla F(\boldsymbol{\lambda}) = \sum^{N}_{a=1}{\frac{1}{Z}\frac{\partial Z_a}{\partial \boldsymbol{\lambda}}} = \sum^{N}_{a=1}{p_a(\bx-\bx_a)}
\label{14}
\eea

\bea 
\bJ(\bx, \boldsymbol{\lambda}) \ = \ \nabla^2 F(\boldsymbol{\lambda}) & & = \ \sum^{N}_{a=1}{\left.\frac{\partial p_a}{\partial \boldsymbol{\lambda}}\right|_\bx}\otimes (\bx-\bx_a)\nn \\
& & = \ \sum^{N}_{a=1}{p_a(\bx-\bx_a)\otimes (\bx-\bx_a)-\br \otimes \br}
\label{15}
\eea

\subsection{\label{pder}Derivatives of the LME shape functions}

The analysis carried out in  \citep{Arrojo:2006} leads  to the following expression of the spatial gradient of $p_a^*$

\bea 
\nabla p^{*}_{a} \ = \ -p^{*}_{a} (\bJ^*)^{-1} (\bx-\bx_a)
\label{A6}
\eea

\noindent where $\bJ^* = \bJ(\bx,  \boldsymbol{\lambda}^*)$.

{\cb{Now we compute the second-order derivatives of the LME shape functions. Differentiating both sides of (\ref{A6}) once, we get}}

\bea 
p^{*}_{a,ij} & & = \ \frac{\partial^2 p^{*}_{a}}{\partial x_i \partial x_j} \ = \ \frac{\partial}{\partial x_j} \left[-p^{*}_{a} J^{*-1}_{ik} (x_k-x_{a_k})\right] \nn \\
& & \nn \\
& & = \ p^{*}_{a,j} J^{*-1}_{ik} (x_k-x_{a_k}) - p^{*}_{a} J^{*-1}_{ij} - p^{*}_{a} J^{*-1}_{ik,j} (x_k-x_{a_k}).
\label{A*}
\eea

The only quantity that needs to be computed on the right-hand side of (\ref{A*}) is $J^{*-1}_{ik,j}$. 
{\cb{Differentiating both sides of the identity shown below}}

\bea 
J^{*}_{im}J^{*-1}_{mj} \ = \ \delta^{i}_{j},
\label{A**}
\eea
 
\noindent where $\delta^{i}_{j}$ denotes the Kronecker symbol, leads to

\bea 
J^{*-1}_{ik,j} \ = \ - J^{*-1}_{im} J^{*}_{mn,j} J^{*-1}_{nk}.
\label{A***}
\eea

\noindent On the other hand, from (\ref{15}) we deduce the result

\bea 
J^{*}_{mn,j} & = & \sum^{N}_{a=1}{p^{*}_{a}\left[\delta^{m}_{j}(x_n-x_{a_n}) + \delta^{n}_{j}(x_m-x_{a_m})\right]} \nn \\
& & \nn \\
& & -\left[\sum^{N}_{a=1}{p^{*}_{a,j}(x_m-x_{a_m}) + p^{*}_{a}\delta^{m}_{j}}\right] r^{*}_{j} \nn \\
& & \nn \\
& & -r^{*}_{m} \left[\sum^{N}_{a=1}{p^{*}_{a,j}(x_n-x_{a_n}) + p^{*}_{a}\delta^{n}_{j}}\right]
\label{A****}
\eea
\noindent Equations (\ref{A***}) {\cb{ and  }}(\ref{A****}) allow us to {\cb{derive the}} explicit formulae (\ref{A*}).


\subsection{\label{curv}Lines of curvature and principal curvatures of membrane networks}

\begin{figure}[htbp]
  \begin{center}
    \setlength{\unitlength}{1mm}
    \epsfig{file=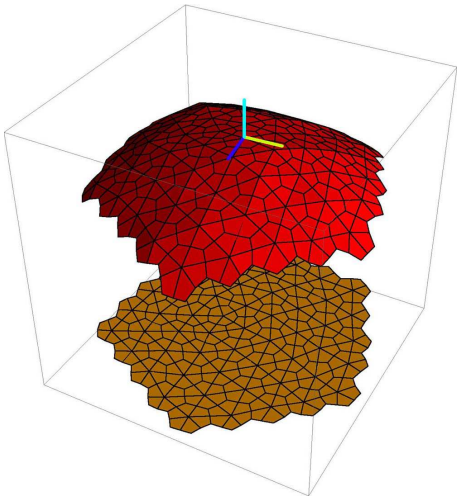,angle=0,width=60mm} 
    \vspace{0.5cm}
   \caption{(Color online) Node set extracted from a 3D membrane network and its orthogonal projection onto the $x_1,x_2$ plane.}
    \label{network}
  \end{center}
\end{figure}

Let us now examine a node set 
${ X}_N =\{ \{ x_{a_1}, x_{a_2}, z_{a} \}, \ a=1,...,N\}$
extracted from a membrane network lying in the 3D Euclidean space (Fig. \ref{network}). 
The Monge chart  

\bea 
z_N(\bx) \ = \ \sum^{N}_{a=1}{ z_{a} p_{a}^*(\bx)}
\label{z_maxent}
\eea

\noindent defines the LME regularization of ${ X}_N$ that we will denote by  ${ S}_N$ in the following. 
The unit vectors 
${\boldsymbol\nu}_{\left\langle 1 \right\rangle}$, ${\boldsymbol\nu}_{\left\langle 2 \right\rangle}$ {\cb{are}}
tangent to the lines of curvature of ${ S}_N$, 
and the principal curvatures $k_1,  k_2$ of such a surface correspond to the solution of the eigenvalue problem (see, e.g., \citep{Stoker:1969}, \citep{Naghdi72}; Appendix A.2)

\bea 
\left(b_{\alpha\beta} \ -\ k_{\gamma} \ a_{\alpha\beta}\right) \ \nu^{\beta}_{(\gamma)} \ = \ 0
\ \ \ (\gamma=1,2)
\label{*}
\eea

\noindent where $a_{\alpha \beta}$ and $b_{\alpha \beta}$ are the first and the second fundamental forms of ${ S}_N$, defined by 

\beq
\label{ab}
\begin{array}{lllll}
a_{\alpha \beta} & = &  \delta^{\alpha}_{\beta} + {z_N},_{\alpha\beta}, \ \ \ \ \ \
b_{\alpha \beta} & = &   -{z_N},_{\alpha\beta} / \sqrt{1+{z_N},^{2}_{1}+{z_N},^{2}_{2}}.
\end{array}
\eeq

By taking into account the previous expressions of  the shape functions  $p_{a}^*$ and their derivatives, we easily derive from Eqns. (\ref{*}) and (\ref{ab}) the LME estimates of the mean curvature $H_N^{\bx}=1/2(k_1  + k_2)$, and the Gaussian curvature $K_N^{\bx}=k_1 k_2$ of ${ S}_N$ at the given $\bx$.

\section{\label{numerics}Numerical results}


\subsection{\label{sinus}Sinusoidal membrane}

\begin{figure}[ht]
\unitlength1cm
\begin{picture}(14.5,8)
\if\Images y\put(0,0.5){\psfig{figure=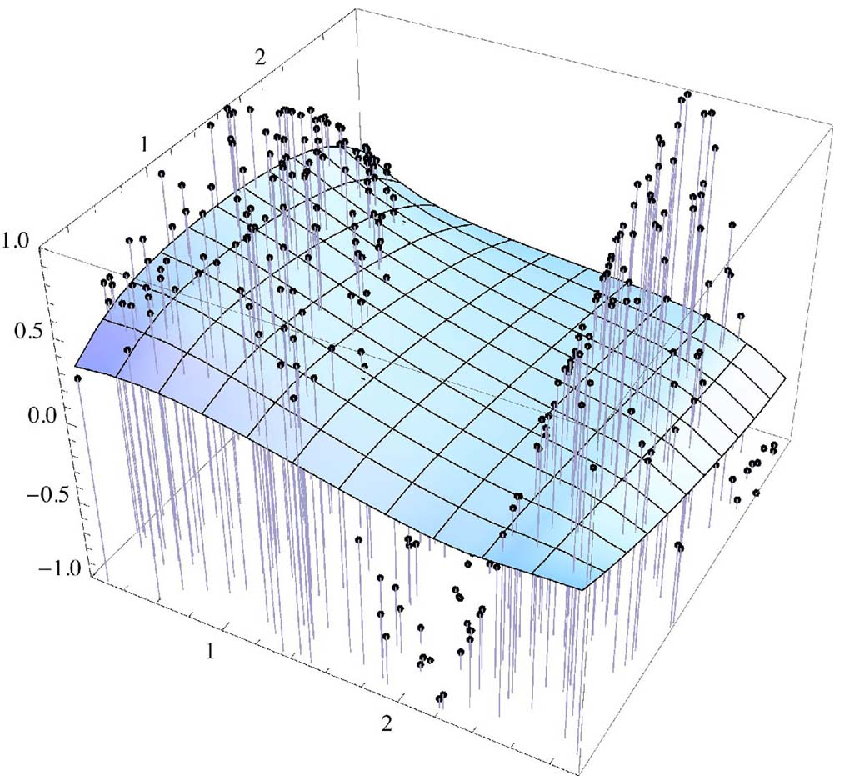,width=7.0cm}}\fi
\if\Images y\put(7.5,0.5){\psfig{figure=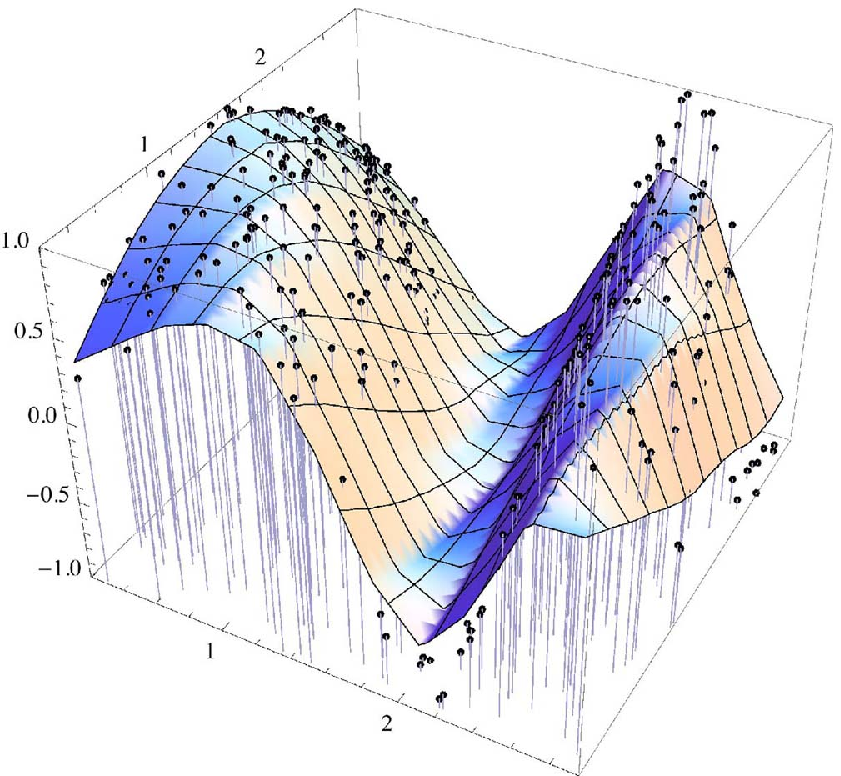,width=7.0cm}}\fi
\if\Images y\put(3,0){$\beta=0.001$}\fi
\if\Images y\put(10.5,0){$\beta=10$}\fi
\end{picture}
\caption{(Color online) LME approximations of scattered data extracted from the  sinusoidal surface
${ S}: \ z=\sin ({x_1}^2+x_2)$ for two different values of $\beta$.}
\label{fig_sinusoidal}
\end{figure}

We begin by considering the node set ${ X}_N$ such that 
$x_{a_1}$ and $x_{a_2}$ are randomly generated numbers within the interval $[0, \pi]$, 
and it results in $z_a=\sin (x_{a_1}^2+x_{a_2})$. Fig. \ref{fig_sinusoidal} shows the LME surfaces ${ S}_N$ obtained in this case for  $N=500$,  $\beta=0.001$, and  $\beta=10$. The ${ S}_N$ are sampled over a $12 \times 12$ uniform grid of points defined over the  $x_1,x_2$ region $[0,\pi] \times [0, \pi]$.
We observe from Fig. \ref{fig_sinusoidal}  that the ${ S}_N$ corresponding to $\beta =0.001$
is almost flat, while the ${ S}_N$ corresponding to $\beta=10$ fairly reproduces the local shape of ${ X}_N$ in the neighborhood of each node.

Next, we examine uniform grids of nodes $\bx_a = \{x_{a_1}, x_{a_2} \}$ over the  $x_1,x_2$ region $D=[0,3] \times [0, 3]$, and the 3D node set
${ X}_N =\{ \hat{\bx}_a=\{ \bx_a, z_a = \sin (x_{a_1}^2+x_{a_2}) \}, \ a=1,...,N\}$. 
For each $\bx_a \in D'=[0.5,2.5] \times [0.5, 2.5] \subset D$, we further consider the subset ${ X}_N^{a} \subset { X}_N$, which is generated by the $m$th nearest neighbors of $\bx_a$, $m$ being an integer parameter. We employ  the node set ${ X}_N^{a}$ to get LME estimates $H_N^{a}$ and $K_N^{a}$ of the mean and Gaussian curvatures of ${ X}_N$ at $\hat{\bx}_a$. 
In order to normalize the effects of $\beta$ on the LME estimates, we rescale such a parameter as follows

\bea 
\beta \ = \ 
\frac{{\bar \beta}}{\left( \mbox{diam}({ X}_N^{a}) \right)^2}
\label{betabar}
\eea

\noindent where $\bar \beta$ is a dimensionless quantity and it results in

\bea 
\mbox{diam}({ X}_N^{a}) \ = \ 
\max_{\bx_a, \bx_b \in { X}_N^{a}}
\{ | \bx_a - \bx_b | \}.
\label{diam}
\eea

\begin{figure}[ht]
\unitlength1cm
\begin{picture}(14.5,4.5)
\if\Images y\put(0.0,0.0){\epsfig{figure=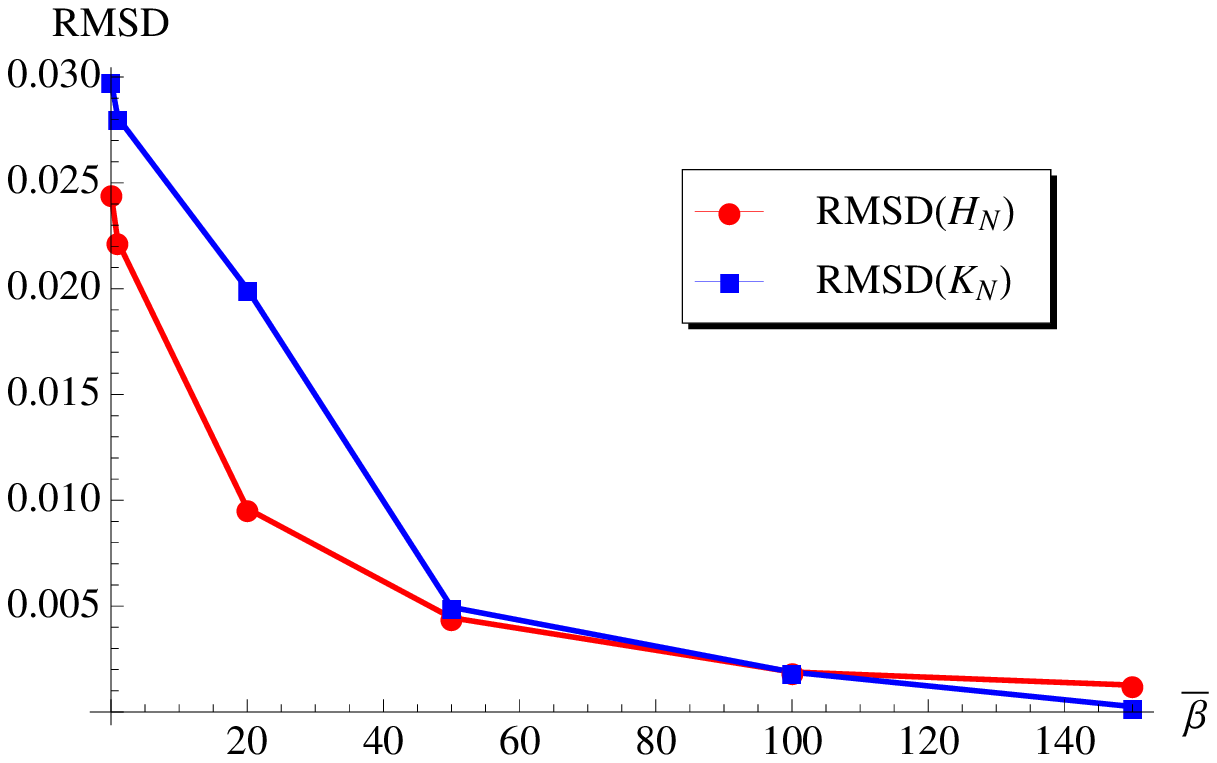,width=7cm}}\fi
\if\Images y\put(7.0,0.0){\epsfig{figure=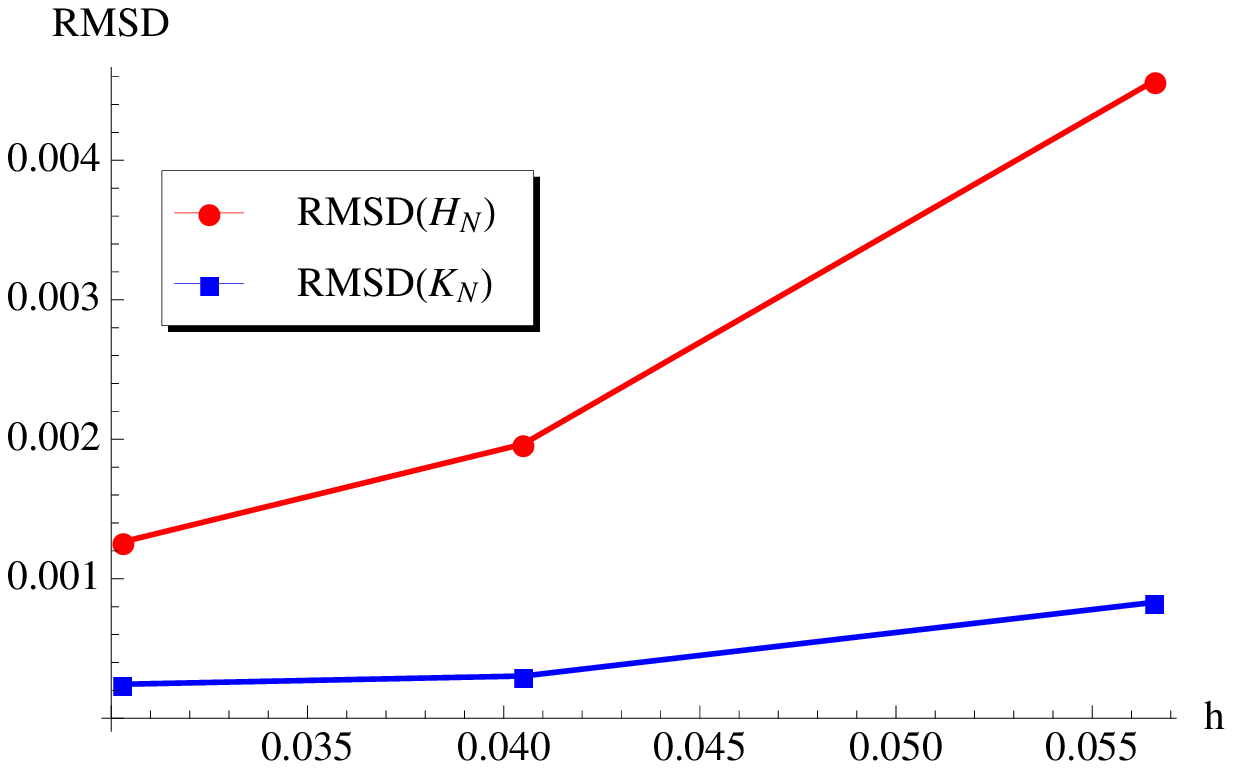,width=7cm}}\fi
\end{picture}
\caption{(Color online) Root Mean Square Deviations (RMSD) of the LME approximations to the mean curvature $H$ and the Gaussian curvature $K$ of the sinusoidal surface $z=sin(x_1^2+x_2)$ over the $x_1,x_2$ domain $[0.5,2.5]\times[0.5,2.5]$. Left: RMSD as a function of $\bar{\beta
}$, for $m=10$ and $h=0.0303$. Right: RMSD as a function of $h$, for $m=10$ and $\bar{\beta
}=150$.}
\label{RMSDsinus}
\end{figure}

\begin{figure}[ht]
\unitlength1cm
\begin{picture}(14.0,15.5)
\if\Images y\put(0,10.66){\epsfig{figure=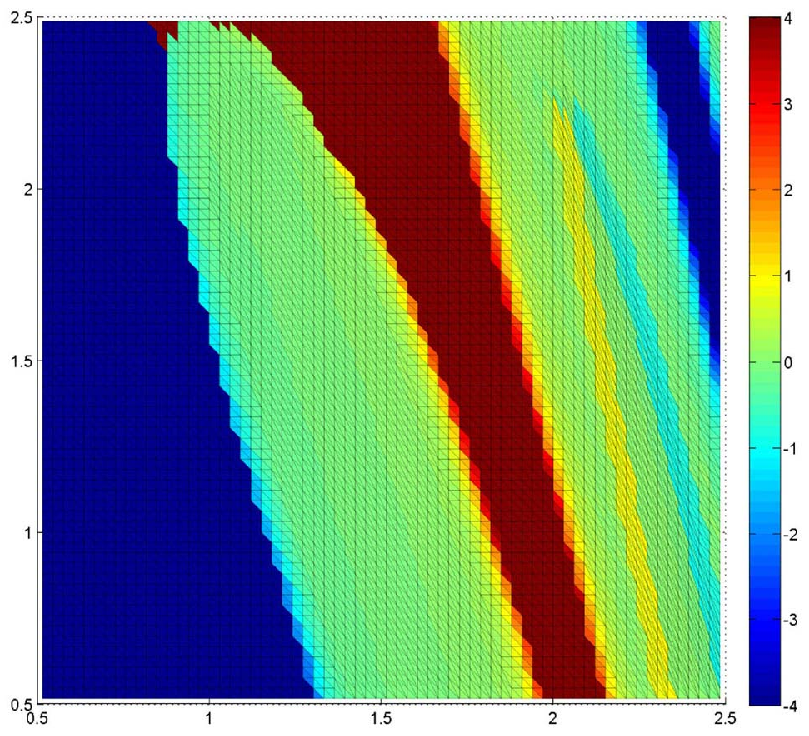, width=5.3cm}}\fi
\if\Images y\put(7,10.66){\epsfig{figure=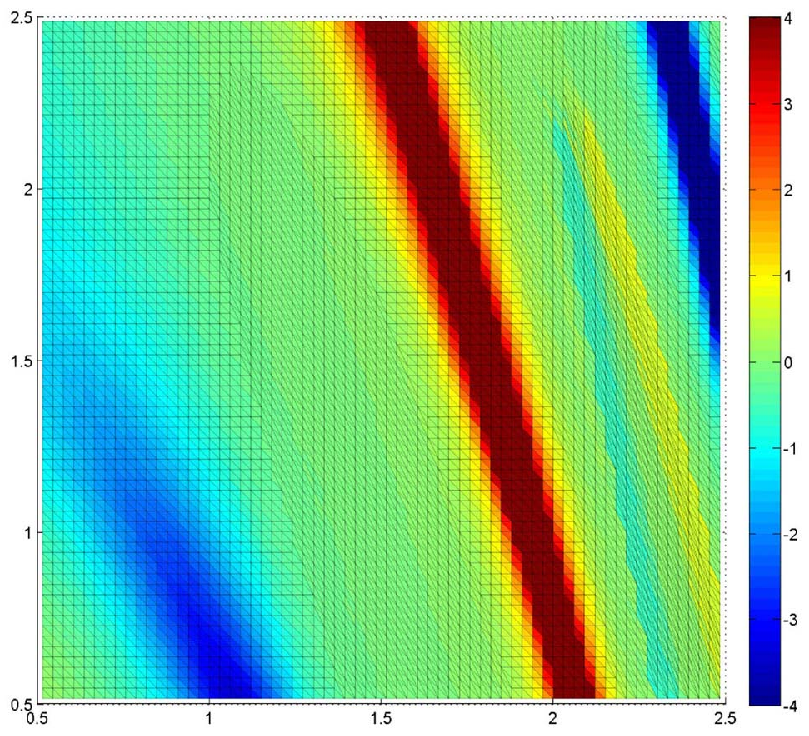, width= 5.3cm}}\fi
\if\Images y\put(0,5.33){\epsfig{figure=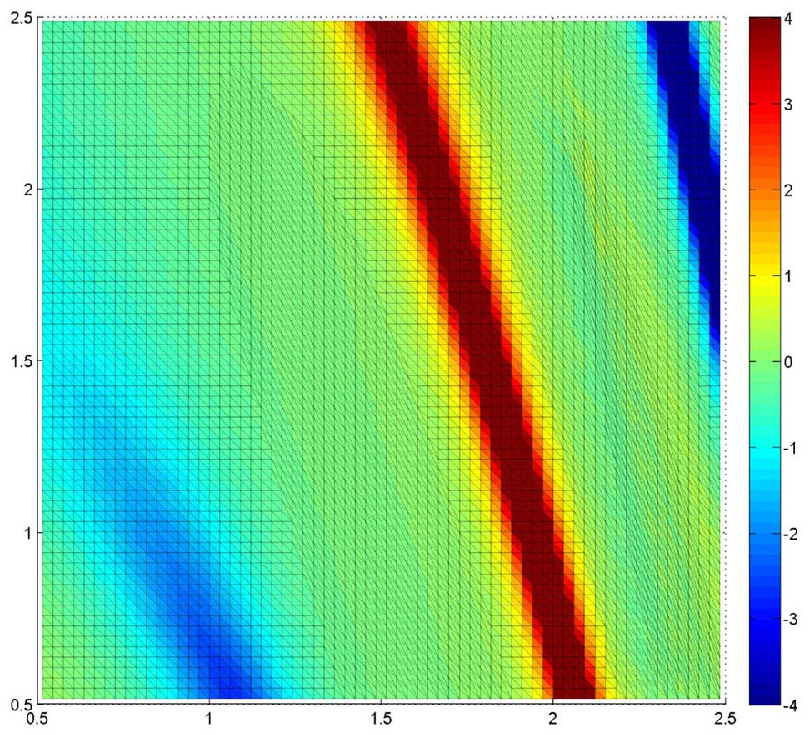, width= 5.3cm}}\fi
\if\Images y\put(7,5.33){\epsfig{figure=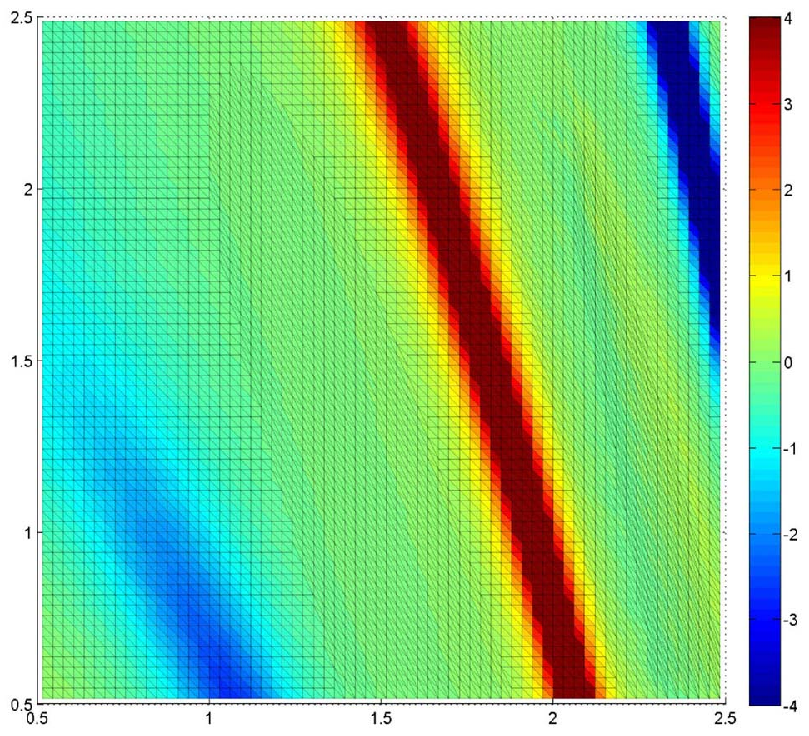,width= 5.3cm}}\fi
\if\Images y\put(0.0,0.0){\epsfig{figure=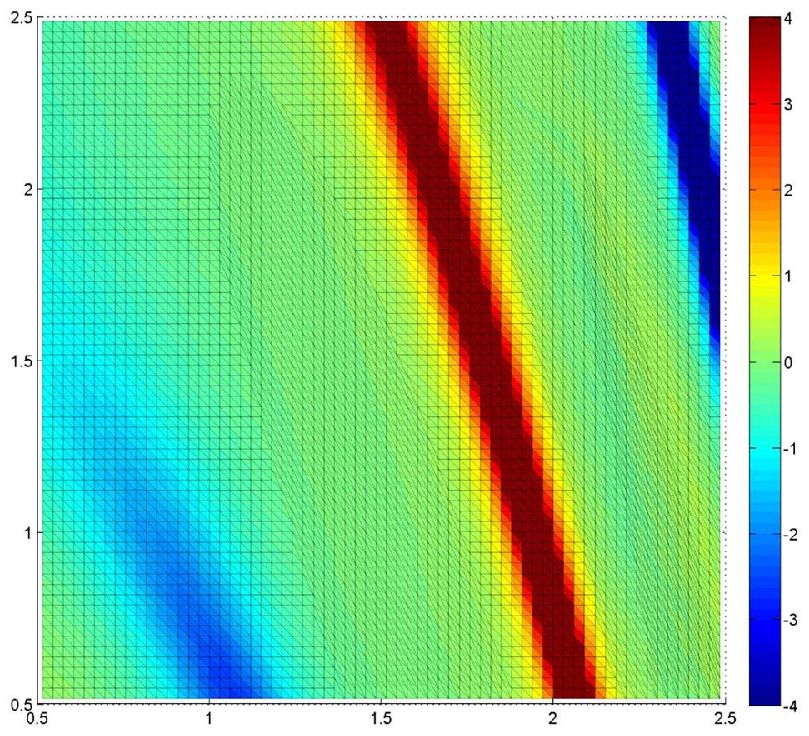,width= 5.3cm}}\fi
\if\Images y\put(7.0,0.0){\epsfig{figure=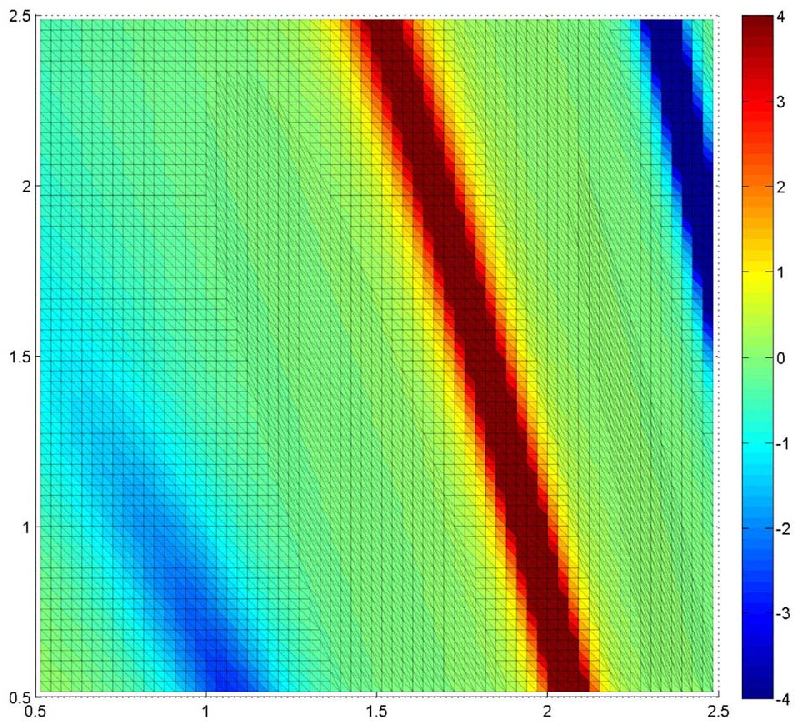,width= 5.3cm}}\fi
\if\Images y\put(0.5,10.55){\footnotesize{$m=5, \ RMSD=9.733\times10^{1}$}}\fi
\if\Images y\put(7.5,10.55){\footnotesize{$m=7, \ RMSD=6.161\times10^{-3}$}}\fi
\if\Images y\put(0.5,5.2){\footnotesize{$m=9, \ RMSD=1.173\times10^{-3}$}}\fi
\if\Images y\put(7.5,5.2){\footnotesize{$m=10, \ RMSD=1.265\times10^{-3}$}}\fi
\if\Images y\put(0.5,-0.15){\footnotesize{$m=12, \ RMSD=1.333\times10^{-3}$}}\fi
\if\Images y\put(7.5,-0.15){\footnotesize{Exact}}\fi
\end{picture}
\caption{(Color online) 2D density plots 
of the LME approximations to the mean curvature $H$ of the sinusoidal surface $z=sin(x_1^2+x_2)$ over the $x_1,x_2$ domain $[0.5,2.5]\times[0.5,2.5]$, for $\bar{\beta
}=150$, $h=0.0303$, and different values  of $m$ {(lower bound of the color bar: -4; .upper bound: +4)}.}
\label{Hsinus}
\end{figure}

\begin{figure}[ht]
\unitlength1cm
\begin{picture}(14.0,16.0)
\if\Images y\put(0,10.66){\epsfig{figure=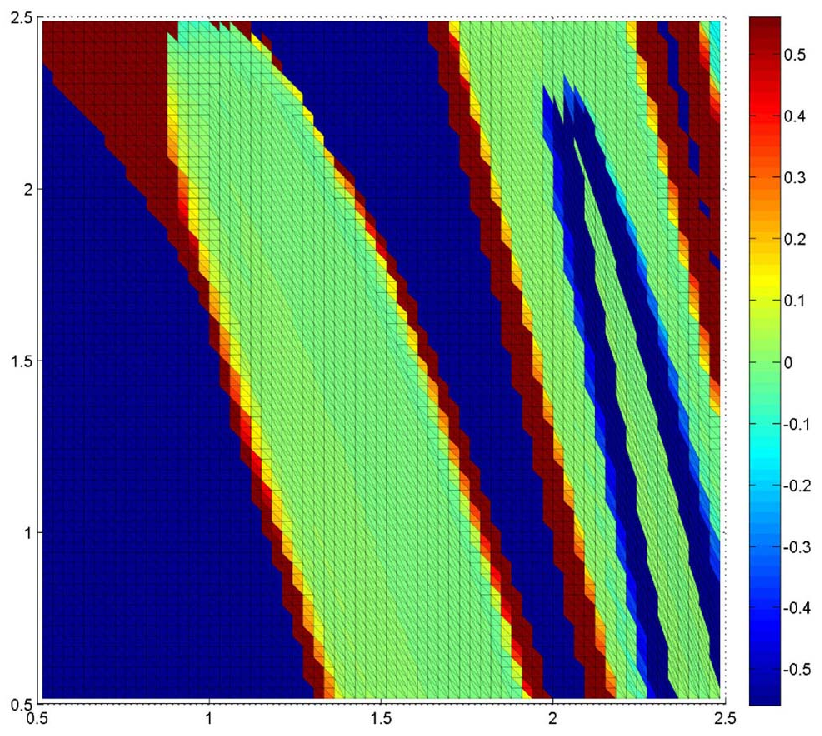,width=5.3cm}}\fi
\if\Images y\put(7,10.66){\epsfig{figure=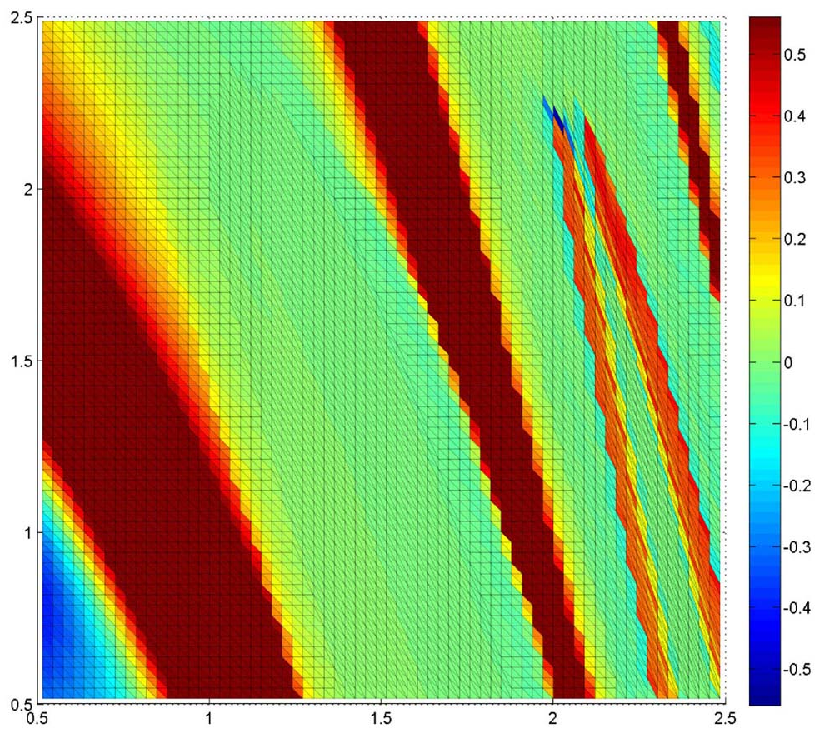,width=5.3cm}}\fi
\if\Images y\put(0,5.33){\epsfig{figure=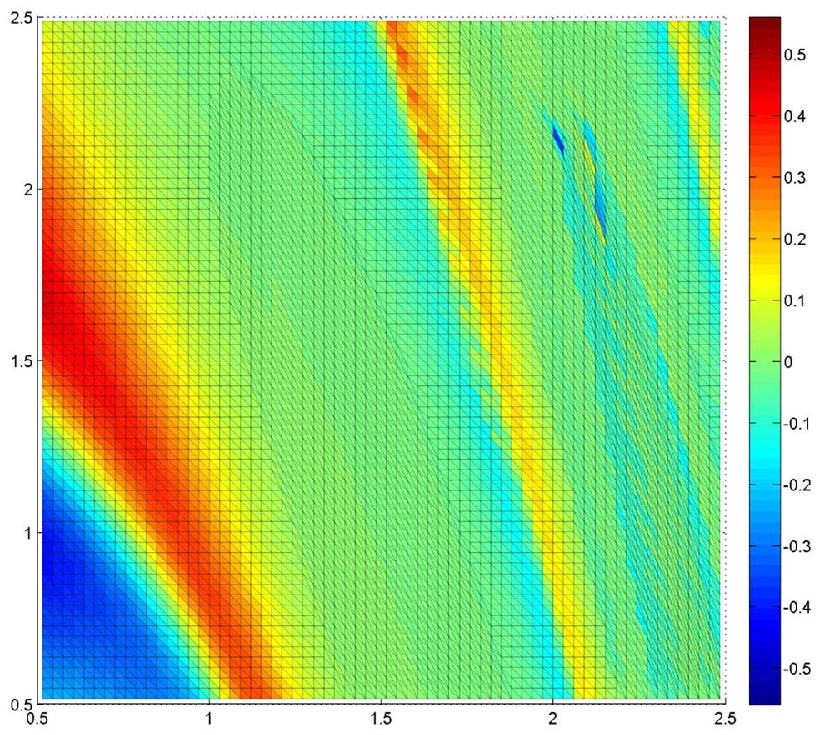,width=5.3cm}}\fi
\if\Images y\put(7,5.33){\epsfig{figure=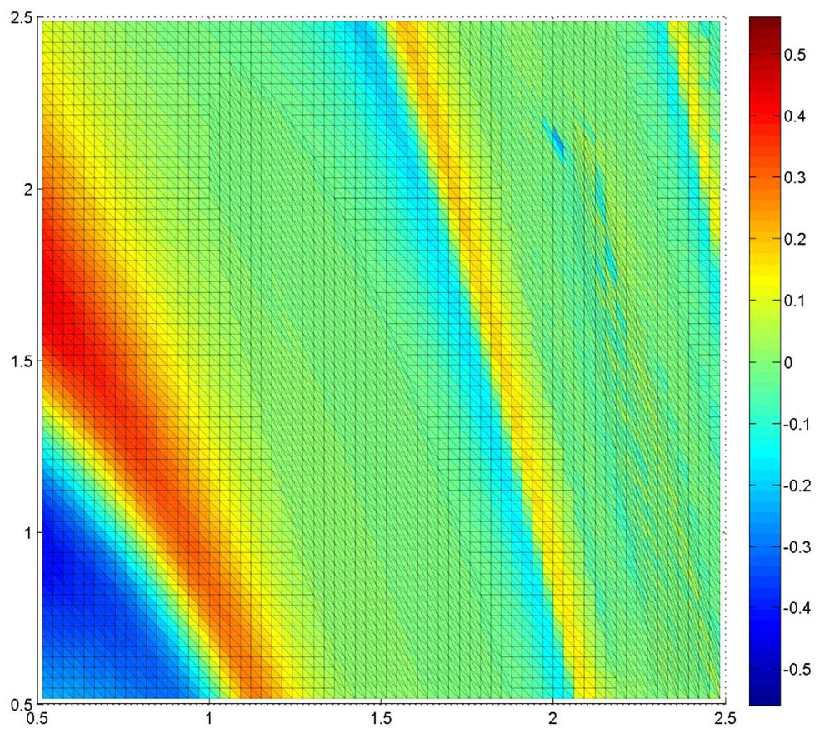,width=5.3cm}}\fi
\if\Images y\put(0.0,0.0){\epsfig{figure=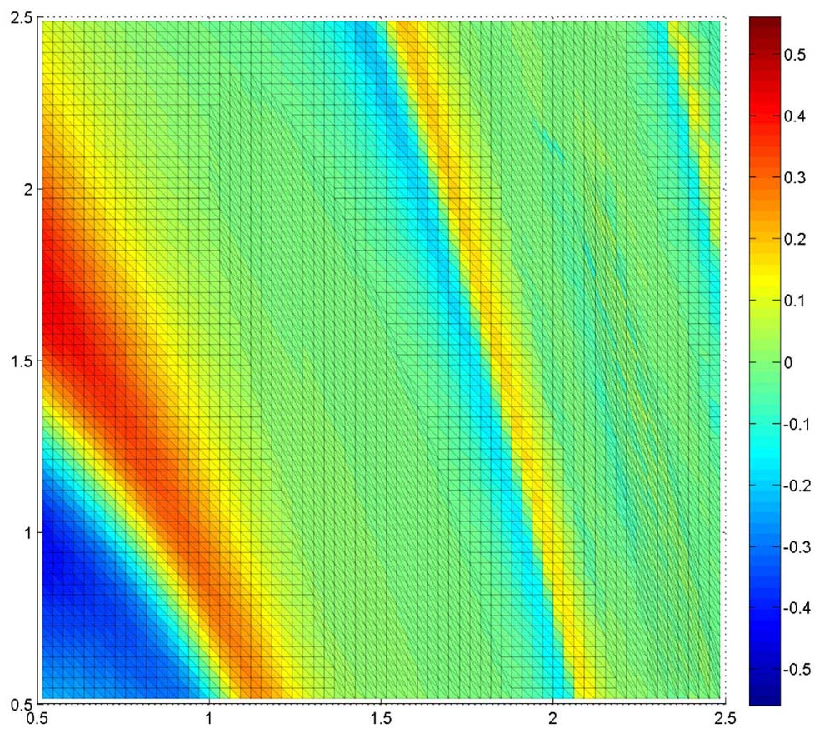,width=5.3cm}}\fi
\if\Images y\put(7.0,0.0){\epsfig{figure=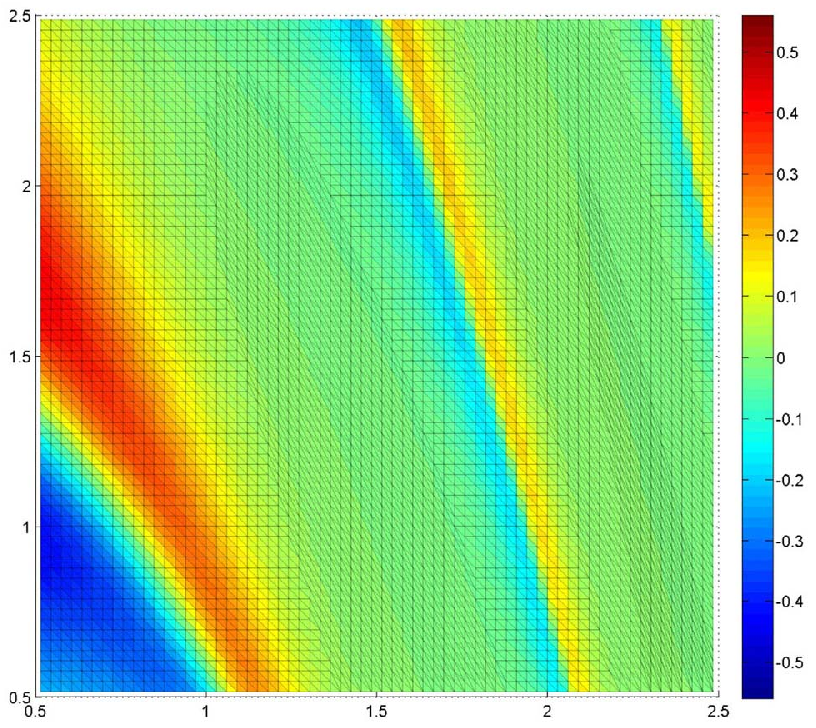,width=5.3cm}}\fi
\if\Images y\put(0.5,10.55){\footnotesize{$m=5, \ RMSD=1.070\times10^{6}$}}\fi
\if\Images y\put(7.5,10.55){\footnotesize{$m=7, \ RMSD=2.089\times10^{-2}$}}\fi
\if\Images y\put(0.5,5.2){\footnotesize{$m=9, \ RMSD=5.730\times10^{-4}$}}\fi
\if\Images y\put(7.5,5.2){\footnotesize{$m=10, \ RMSD=2.438\times10^{-4}$}}\fi
\if\Images y\put(0.5,-0.15){\footnotesize{$m=12, \ RMSD=2.328\times10^{-4}$}}\fi
\if\Images y\put(7.5,-0.15){\footnotesize{Exact}}\fi
\end{picture}
\caption{(Color online) 2D density plots 
of the LME approximations to the Gaussian curvature $K$ of the sinusoidal surface $z=sin(x_1^2+x_2)$ over the $x_1,x_2$ domain $[0.5,2.5]\times[0.5,2.5]$, for $\bar{\beta
}=150$, $h=0.0303$, and different values  of $m$ {(lower bound of the color bar: -0.56; upper bound: +0.56)}.}
\label{Ksinus}
\end{figure}

\begin{figure}[ht]
\unitlength1cm
\begin{picture}(14.5,13.0)
\if\Images y\put(0,0.5){\epsfig{figure=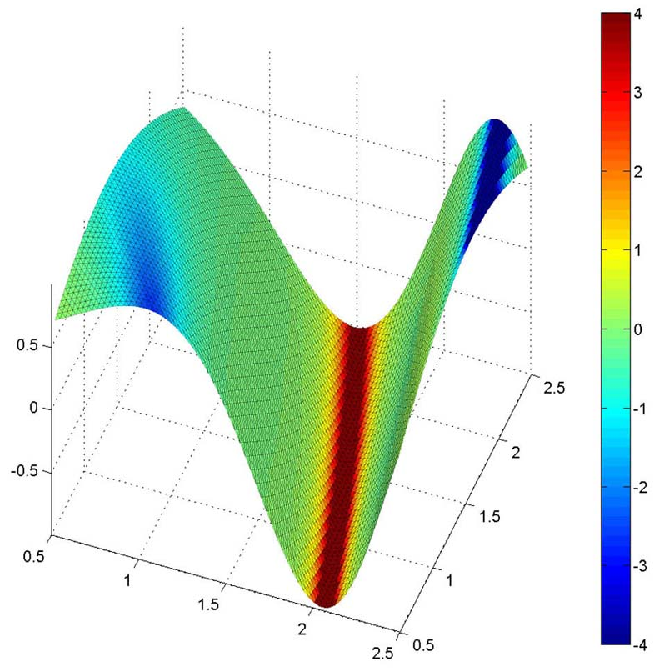,height=7.0cm}}\fi
\if\Images y\put(7.0,0.5){\epsfig{figure=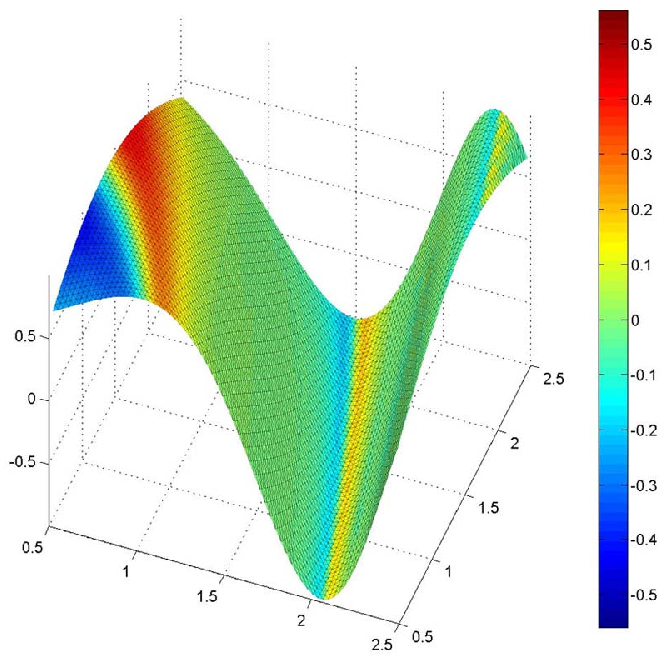,height=7.0cm}}\fi
\if\Images y\put(3.25,0.45){${ H}_N$}\fi
\if\Images y\put(10.25,0.45){${ K}_N$}\fi
\end{picture}
\caption{(Color online) 3D density plots 
of the LME approximations to the mean and the Gaussian curvatures of the sinusoidal surface $z=sin(x_1^2+x_2)$ over the $x_1,x_2$ domain $[0.5,2.5]\times[0.5,2.5]$, for $\bar{\beta
}=150$, $h=0.0303$, and $m=12$.}
\label{HKsinus3D}
\end{figure}

It is useful to compare $H_N^{a}$ and $K_N^{a}$ with the `exact' counterparts $H^a$ and $K^a$, which are
easily computed through (\ref{*}) and (\ref{ab}), by replacing $z_N$ with $z=\sin ({x_1}^2+x_2)$. The accuracy of the nodal LME estimates ${ H}_N=\{H_N^{a}, \ a=1,...,N'\}$  and
${ K}_N=\{K_N^{a}, \ a=1,...,N'\}$ can be inspected by examining the following Root Mean Square Deviations (RMSD)

\beq
\label{RMSD}
\begin{array}{lll}
RMSD({ H}_N) & = &  \sqrt{ \left( \sum^{N'}_{a=1}{(H_N^{a}-H^a)^2} \right) / N'}, \\
RMSD({ K}_N) & = &  \sqrt{ \left( \sum^{N'}_{a=1}{(K_N^{a}-K^a)^2} \right) / N'}
\end{array}
\eeq

\noindent for different values of $\bar{\beta}$ and $m$. Here, $N'$ denotes the total number of nodes belonging to $D'$. 
We examine the following three different mesh sizes: $h=0.0566$ ($N=54\times54$); 
$h=0.0405$ ($N=75\times75$);
and $h=0.0303$ ($N=100\times100$).
Here and in the following examples, we solve the nonlinear optimization problem (\ref{11}) {\cb{using}} recursive Newton-Raphson updates (\ref{13}), up to the termination condition $| \br^k | < 10^{-6} \ \mbox{diam}({ X}_N^{a})$.
Fig. \ref{RMSDsinus} illustrates how the quantities $RMSD({ H}_N)$ and  $RMSD({ K}_N)$ vary with $\bar{\beta}$ and $h$ for fixed $m=10$, while Figs. \ref{Hsinus}, \ref{Ksinus} and \ref{HKsinus3D} depict 2D and 3D density plots of the data sets ${ H}_N$  and ${ K}_N$ for several values of $m$,  keeping $\bar{\beta}=150$, and $h=0.0303$ fixed. 
The results shown in {\cb{Figures \ref{Hsinus}, \ref{Ksinus} and \ref{HKsinus3D}}} point out that the LME estimates ${ H}_N$ and ${ K}_N$ exhibit uniform asymptotic convergence to the exact solutions $H$ and $K$, respectively, for ${\bar \beta} \ge 100$, $m \ge 9$, and $h \le 0.0303$.

\subsection{\label{sphere}Spherical membrane}

We examine, in the present and the following 
sections, a closed membrane network  ${ X}_N$ corresponding to the CGMD model 
of the red blood cell given in \citep{Marcelli:2005, Hale:2009}.
Such a model
describes the system formed by the cytoskeleton
spectrin network, the lipid bilayer and the transmembrane proteins of an actual RBC membrane, through a network of $N$ (virtual) particles embedded in a closed polyhedral surface showing $M$
triangular facets. 
{\cred{
Each particle has sixfold coordination}}, {\cb{ with the exception of twelve `defects,' which instead have fivefold coordination.}}
The particles represent discrete areas of the RBC membrane and their
equilibrium distance $r_0$ is set equal to the average length of the spectrin filaments ($\sim$ 100 nm). Each particle has mass $m$ and is connected to {\cb{it's}} nearest neighbors though linear
springs of {\cb{stiffness}} $k$, which {\cb{are parameterized such that the network of particles reproduces the membrane rigidity of a RBC}}. {\cb{The bending}} rigidity is
also accounted for by {\cb{ introducing dihedral angle potential energy terms}} of angular stiffness $D$ between
adjacent triangles. The global surface area of the polyhedral membrane is kept constant using
Lagrange multipliers, in order to mimic the relative incompressibility of the lipid bilayer of a normal RBC. The model under consideration is also able to keep {\cb{the volume enveloped by the polyhedral membrane constant}}, with the aim to resemble the typical behaviour of a RBC in normal conditions. 
{\cred{
The dynamics of the RBC model can be obtained by integrating the Newton equations of motion of each virtual particle using a standard MD code. We used {\cb{DL POLY 2.20}}(\citep{Smith:1999}), and employed the Nos\'e-Hoover thermostat;
{\cb{$6 \times10^{6}$}} steps with time-step $\Delta t= 2.07 \times 10^{-5} \ t_0$, where $t_0=\sqrt{{m}/{k}}$;
$N=5762$ particles;
$m = 5.82625 \times 10^{-20}$ kg;
$k  = 8.3 \  \mu$N/m; 
$D = 130 \times 10^{-20}$ J; 
constant absolute temperature $T=309$ K; 
{\cc{
constant surface area $\bar{A}=4.986 \times 10^{7} \ \mbox{nm}^2$;
and constant volume $\bar{V}=3.311 \times 10^{10} \ \mbox{nm}^3$.}}
Such parameter settings allow the LME regularization to approximate a spherical surface ${ S}$ with radius ${\bar r} = 1992$ nm \citep{Marcelli:2005}. 
}}

The theoretical average surface ${ S}$ clearly has
uniform principal curvatures $k_1=k_2=H=-1/{\bar r}=-5.019 \times 10^{-4} \ \mbox{nm}^{-1}$ (the minus sign follows from the outward orientation of the normal vector), and Gaussian curvature $K=k_1k_2=25.19 \times 10^{-8} \ \mbox{nm}^{-2}$.
Nodal LME estimates 
${ H}_N=\{H_N^{a}, \ a=1,...,N\}$  and
${ K}_N=\{K_N^{a}, \ a=1,...,N\}$ of the mean and Gaussian curvatures of the network
can be obtained by introducing different local frames $\{ \hat{\bx}_a, \ x_1, \ x_2, \ z \}$ at each different node $\hat{\bx}_a$, with $x_1$, $x_2$ and $z$ tangent to the local parallel, meridian, and radial lines, respectively.
Table \ref{tab:sphere_HK}  shows the mean values and the standard deviations of  the data sets ${ H}_N$ and ${ K}_N$ defined as follows

\beq
\label{Hsphere_statistics}
\begin{array}{llllll}
{\bar { H}}_N & = & 1/N \ \sum^{N}_{a=1}{H_N^a}, \ \ \
s_d({ H}_N) & = &  \sqrt{ \left( \sum^{N}_{a=1}{(H_N^{a}-{\bar H}_N)^2} \right) / (N-1)}
\end{array}
\eeq

\beq
\label{Ksphere_statistics}
\begin{array}{llllll}
{\bar { K}}_N & = & 1/N \ \sum^{N}_{a=1}{K_N^a}, \ \ \
s_d({ K}_N) & = &  \sqrt{ \left( \sum^{N}_{a=1}{(K_N^{a}-{\bar K}_N)^2} \right) / (N-1)}
\end{array}
\eeq

\noindent for $\bar{\beta}=100$ and different values of $m$. The results in Table \ref{tab:sphere_HK} highlight a {\cb{ good agreement}} between LSM estimates and exact solutions for $m \ge 10$. 
{\cc{
It has to be considered that the MD model doesn't reach a perfectly (average) spherical shape at equilibrium, due to the presence of the fivefold defects}}.

\begin{table}[htbp]
\begin{centering}
\begin{tabular}{|c|c|c|c|c|c|}
\hline 
& $m=5$ &  $m=7$ & $m=9$ & $m=10$ &
$m=12$  \\
\hline 
${\bar { H}}_N \times 10^{5} \ \mbox{\AA}^{-1}$ & 39.0626 & -7.1062  & -5.1293 & -5.0443 & -5.0311  \\
\hline 
$s_d({ H}_N) \times 10^{5} \ \mbox{\AA}^{-1}$ & 2.8047 & 0.1467 & 0.0296  & 0.0271 & 0.0238 \\
\hline 
${\bar { K}}_N \times 10^{10} \ \mbox{\AA}^{-2}$ & 1540.19 & 50.488 & 26.3077 & 25.4453 & 25.3118  \\
\hline 
$s_d({ K}_N) \times 10^{10} \ \mbox{\AA}^{-2}$ & 234.04 & 2.1551 & 0.3152  & 0.2840 & 0.2473 \\
\hline 
\end{tabular}
\par\end{centering}
\caption[]{Mean and standard deviation of the data sets ${ H}_N$ and ${ K}_N$ for a spherical membrane network with $N=5762$ nodes and radius $r=1992 \ \mbox{nm}$, considering $\bar{\beta
}=100$ and different values of $m$. Exact solution: $H=-5.019 \times 10^{-5} \ \mbox{\AA}^{-1}$, $K=25.19 \times 10^{-10} \ \mbox{\AA}^{-2}$.}
\label{tab:sphere_HK} 
\end{table}

\bigskip 


\subsection{\label{RBC}Principal curvatures of the RBC membrane}

\begin{figure}[ht]
\unitlength1cm
\begin{picture}(12,7.5)
\if\Images y\put(1,0){\psfig{figure=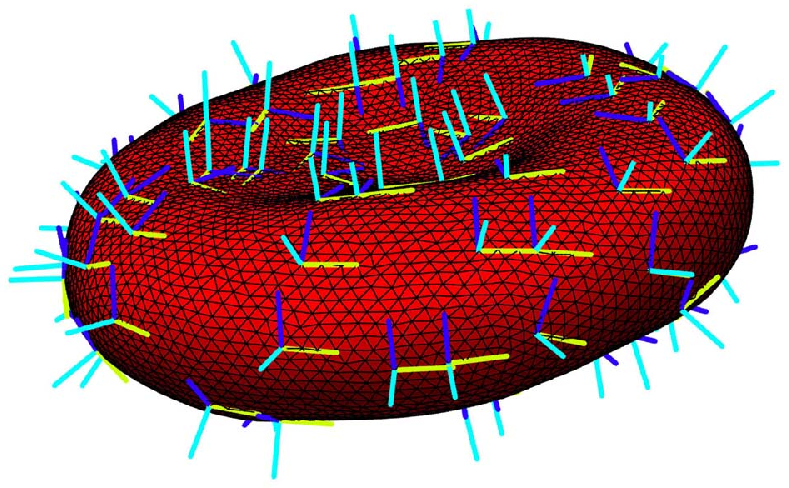,width=12.0cm, angle=0}}\fi
\end{picture}
\caption{{\cred{
(Color online) 3D map of the local bases $\{ \hat{\bx}_a, \ x_1, \ x_2, \ z \}$ introduced at selected nodes of a triangulated model of the RBC membrane ($x_1$: yellow, $x_2$: blue, $z$: cyan)
}}.}
\label{localbasis_RBC}
\end{figure}

We analyze in the present 
section a slight different formulation of the RBC model  given in \citep{Marcelli:2005,Hale:2009}, which differs from that {\cb{discussed}} in the previous section {\cb{only in terms of the}} 
{\cc{
volume constraint. Here, we set the volume enveloped by the RBC membrane equal to 0.65 times the volume of the sphere analyzed in the previous example, 
allowing the current model to assume the typical biconcave shape of a normal RBC  \citep{Hale:2009}.The surface area of the RBC membrane is again set to $4.986 \times 10^{7} \ \mbox{nm}^2$, as in the previous case. }}
We compute nodal LME estimates 
${ H}_N=\{H_N^{a}, \ a=1,...,N\}$  and
${ K}_N=\{K_N^{a}, \ a=1,...,N\}$  of the mean and Gaussian curvatures of a real RBC membrane by processing the rolling average configurations ${X}_N$ of the CGMD model up to different simulation times $t$.
{\cred{
Let $n$ denote the weighted unit normal to the current vertex $\hat{\bx}_a$ of the triangulation associated with ${X}_N$,  assuming the triangle areas as the weights \citep{Taubin:1995}. 
In the present case, we define the local $x_1$ axis as the direction of the edge attached to  $\hat{\bx}_a$ that has the minimum deviation from the parallel drawn on an ideal sphere passing through the same point. In addition, we define $x_2$ by means of the vector product of the unit vectors in the directions of $n$ and $x_1$, and $z$ via the vector product of the unit vectors in the directions of $x_1$ and $x_2$. A graphical representation of the local parameterization introduced for the analyzed model of the RBC membrane is provided in Fig. \ref{localbasis_RBC}.}}

\begin{figure}[ht]
\unitlength1cm
\begin{picture}(14.5,6)
\if\Images y\put(3.0,0.0){\epsfig{figure=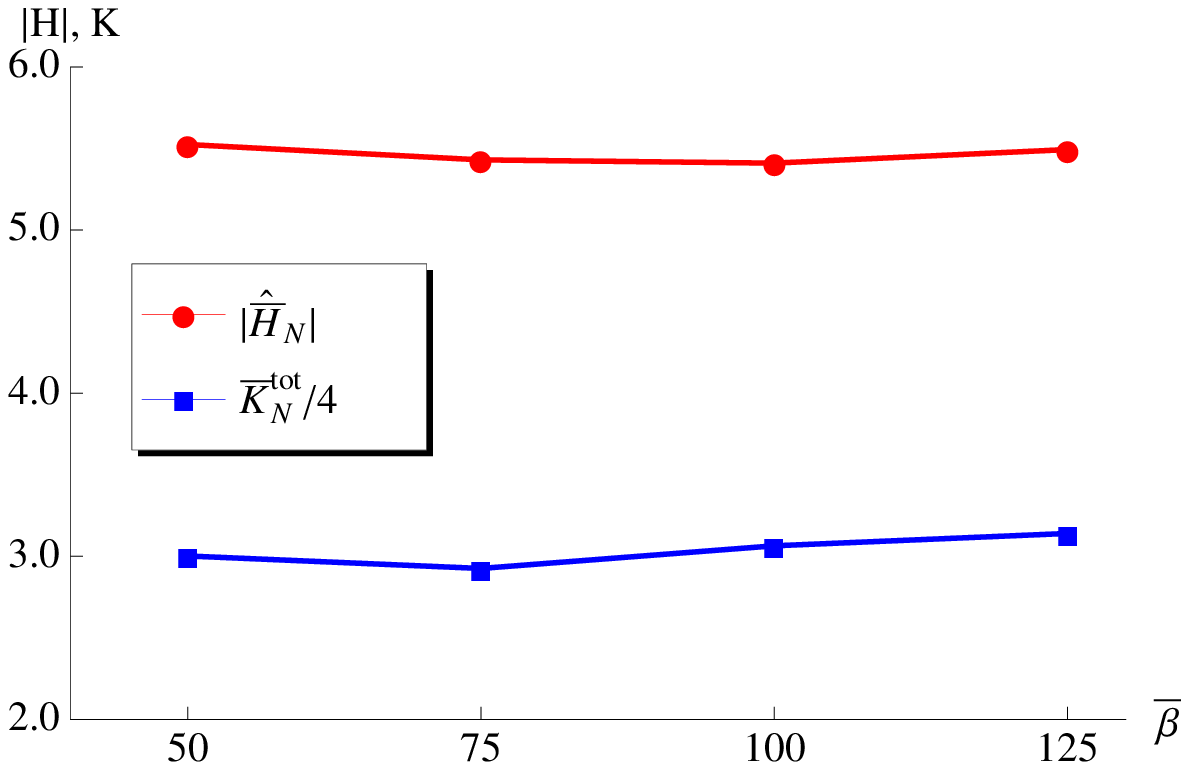,width=8cm}}\fi
\end{picture}
\caption{(Color online) LME estimates  $| {\hat {\bar H}}_N | = | {\bar H}_N | \times 10^{5} \ \mbox{\AA}^{-1}$ and ${ K}_N^{tot}$ 
for a CGMD model of the RBC membrane \cite{Marcelli:2005,Hale:2009},
considering different values of $\bar{\beta}$, $m=10$, and $t=75.5 \ t_0$.}
\label{RBC_HK}
\end{figure}

\begin{figure}[ht]
\unitlength1cm
\begin{picture}(14.0,11.5)
\if\Images y\put(-1,6.33){\epsfig{figure=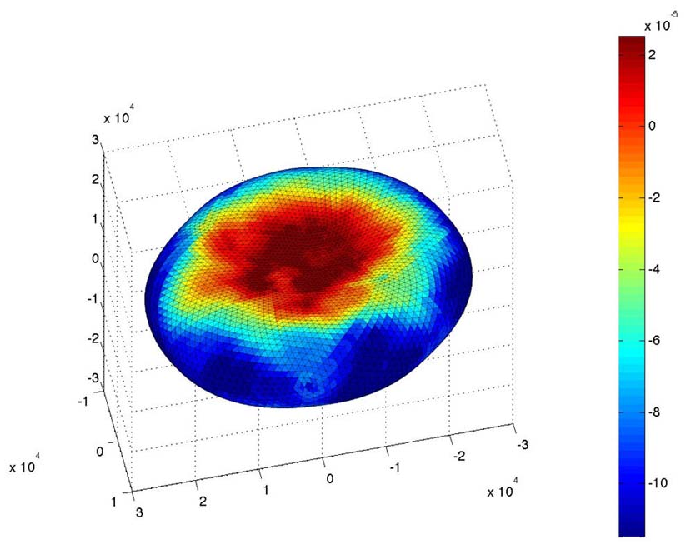,height=5.8cm}}\fi
\if\Images y\put(6,6.33){\epsfig{figure=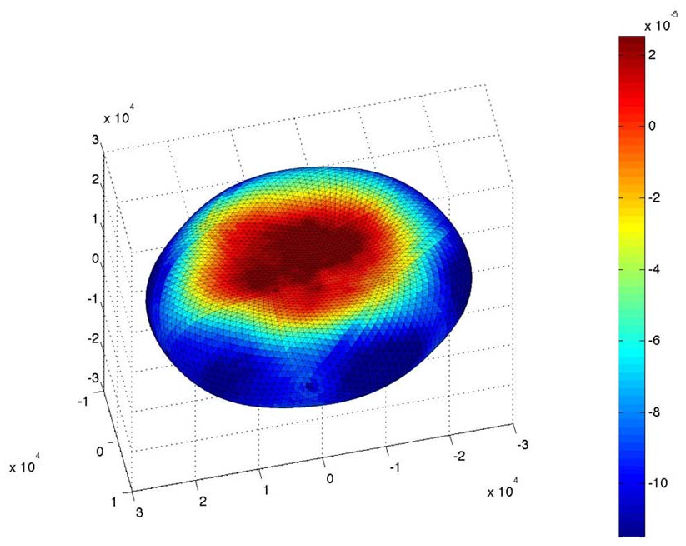,height=5.8cm}}\fi
\if\Images y\put(-1.0,0.5){\epsfig{figure=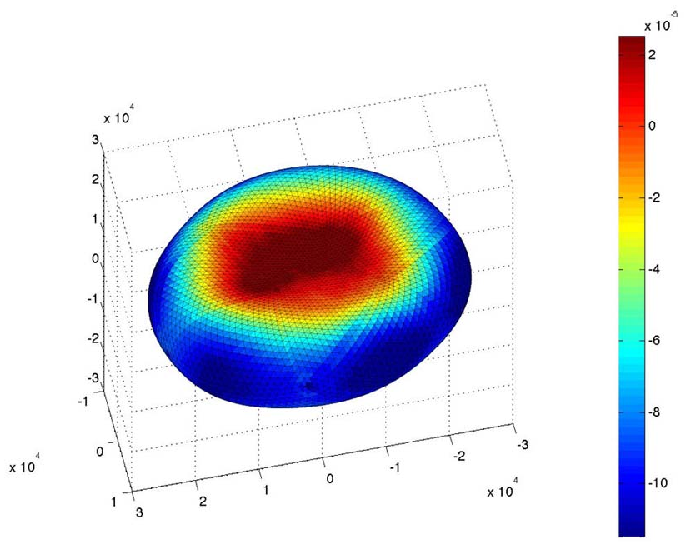,height=5.8cm}}\fi
\if\Images y\put(6.0,0.5){\epsfig{figure=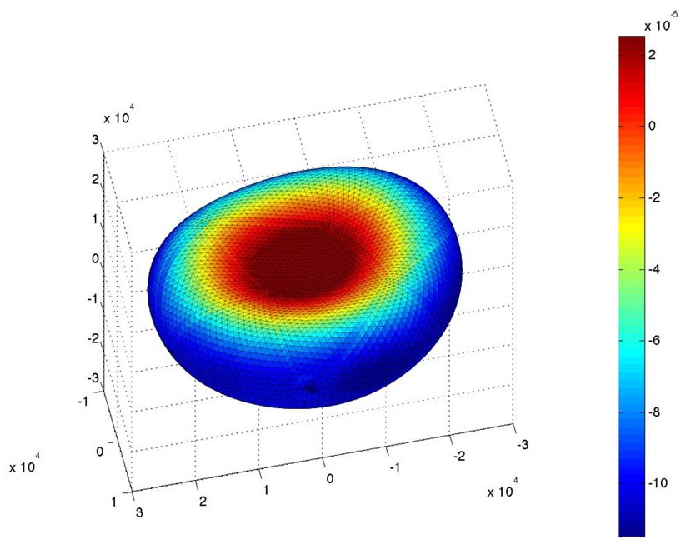,height=5.8cm}}\fi
\if\Images y\put(0.5,5.6){\footnotesize{$t = 2 \ t_0, \ {\hat {\bar H}}_N=-5.61$}}\fi
\if\Images y\put(7.5,5.6){\footnotesize{$t = 5 \ t_0, \ {\hat {\bar H}}_N=-5.62$}}\fi
\if\Images y\put(0.5,0.15){\footnotesize{$t = 10 \ t_0, \ {\hat {\bar H}}_N=-5.61$}}\fi
\if\Images y\put(7.5,0.15){\footnotesize{$t = 71.5 \ t_0, \ {\hat {\bar H}}_N=-5.49$}}\fi
\end{picture}
\caption{
{\cred{
(Color online) 3D density plots 
of the LME approximations ${H}_N$ to the mean curvature $H$ of
of a CGMD model of the RBC membrane 
\cite{Marcelli:2005,Hale:2009}
at different simulation times $t$, 
for $\bar{\beta}=125$ and $m=10$ 
($ {\hat {\bar H}}_N = {\bar H}_N \times 10^{5} \ \mbox{\AA}^{-1}$; lower bound of the color bar: $-11.5 \times 10^{-5} \ \mbox{\AA}^{-1}$; upper bound:  $+2.5 \times 10^{-5} \ \mbox{\AA}^{-1}$).
}}}
\label{HRBC662}
\end{figure}

\begin{figure}[ht]
\unitlength1cm
\begin{picture}(14.0,11.5)
\if\Images y\put(-1,6.33){\epsfig{figure=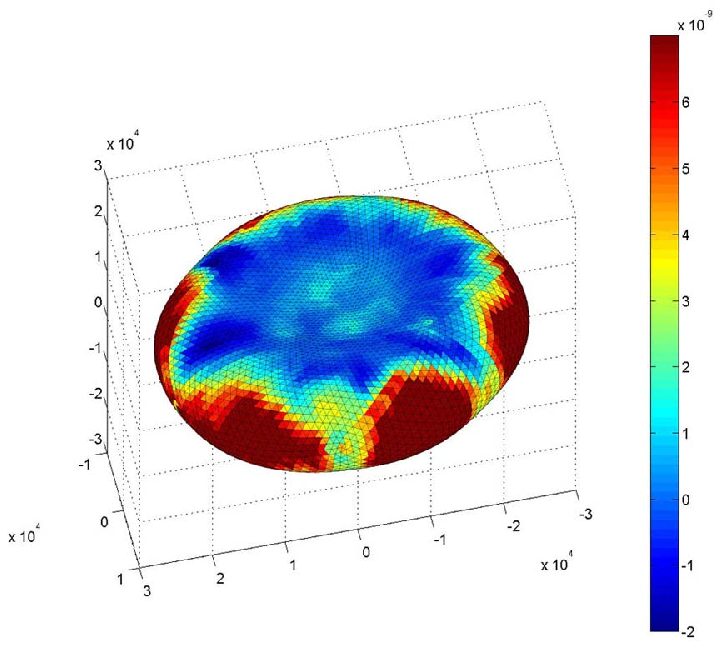,height=6cm}}\fi
\if\Images y\put(6,6.33){\epsfig{figure=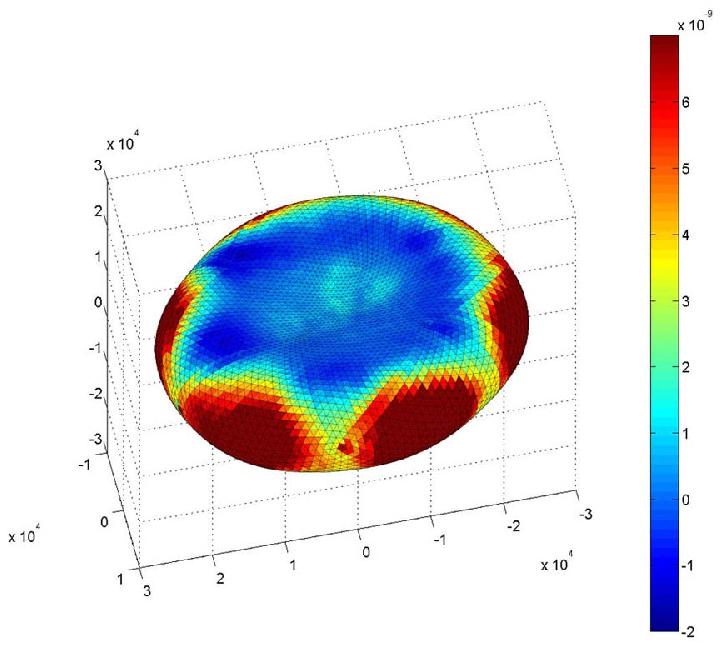,height=6cm}}\fi
\if\Images y\put(-1,0.5){\epsfig{figure=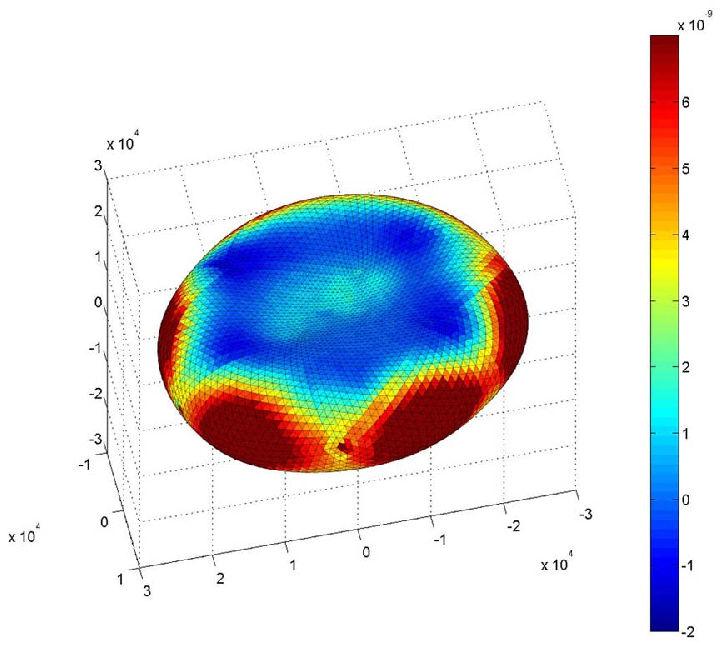,height=6cm}}\fi
\if\Images y\put(6.0,0.5){\epsfig{figure=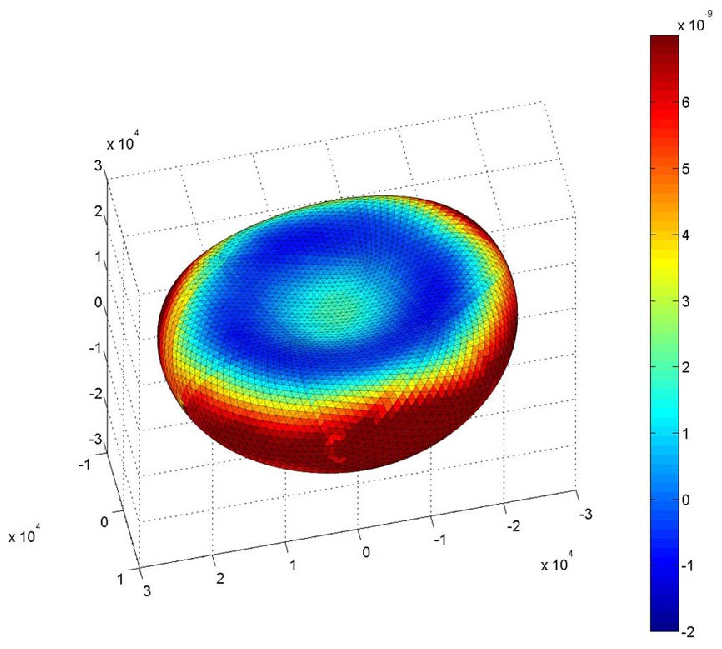,height=6cm}}\fi
\if\Images y\put(0.5,5.6){\footnotesize{$t = 2 \ t_0, \  {K}_N^{tot} /4 = 3.26$}}\fi
\if\Images y\put(7.5,5.6){\footnotesize{$t = 5 \ t_0, \  {K}_N^{tot} /4 = 3.25$}}\fi
\if\Images y\put(0.5,0.15){\footnotesize{$t = 10 \ t_0, \  {K}_N^{tot} /4 = 3.22$}}\fi
\if\Images y\put(7.5,0.15){\footnotesize{$t = 71.5 \ t_0, \  {K}_N^{tot} /4 = 3.14$}}\fi
\end{picture}
\caption{
{\cred{
(Color online) 3D density plots 
of the LME approximations ${K}_N$ to the Gaussian curvature $K$ of
of a CGMD model of the RBC membrane
\cite{Marcelli:2005,Hale:2009}
at different simulation times $t$, 
for $\bar{\beta}=125$ and $m=10$ 
(lower bound of the color bar: $-2 \times 10^{-9} \ \mbox{\AA}^{-2}$; upper bound:  $+7 \times 10^{-9} \ \mbox{\AA}^{-2}$).
}}}
\label{KRBC662}
\end{figure}

Fig. \ref{RBC_HK} shows the dependence of the LME estimates
${\bar H}_N$  and
${K}_N^{tot}  = {\bar K}_N {\bar A}$  on $\bar{\beta}$, for $t = 75.5 \ t_0$  and $m=10$.
It is worth noting that ${K}_N^{tot}$ represents an estimate of the total curvature ${K}^{tot}=\int_{S_N} K \ dA$ of the RBC model.
Due to the `Gauss-Bonet theorem' such a quantity only depends on the genus of $S_N$ and should be equal to $4 \pi$, as in the case of a sphere 
\citep{Stoker:1969}.
One observes from Fig.  \ref{RBC_HK} slight oscillations of ${\bar H}_N$ and ${K}_N^{tot}$ with $\bar{\beta}$; ${K}_N^{tot}/4 \approx 3.00$  for  $\bar{\beta} < 100$; and ${K}_N^{tot}/4 \approx 3.14$ for  $\bar{\beta}=125$. 
3D density plots of the data sets ${ H}_N$  and ${K}_N$ are given in Figs.  \ref{HRBC662} {\cb{and}} \ref{KRBC662}, considering $\bar{\beta}=125$,  $m=10$, and different simulation times $t$. 
The results shown in Figs.  \ref{HRBC662} {\cb{and \ref{KRBC662}}} indicate that the LME regularizations of the average CGMD configurations smoothly describe the geometry of a normal RBC membrane. The LME regularization is indeed able to reproduce the biconcave shape of such a membrane (consider that the hidden bottom edges of the surfaces shown in Figs.  \ref{HRBC662} {\cb{and}} \ref{KRBC662} are nearly specular with respect to the top ones), furnishing  positive (red) and negative (blue) mean curvatures in correspondence with concave and convex portions, respectively (cf. Fig.  \ref{HRBC662}), and negative (blue)  Gaussian curvatures over saddle-shaped regions (cf. Fig.  \ref{KRBC662}).
{\cc{ One observes perturbations in the mean and Gaussian curvature maps in correspondence with the fivefold defects of the membrane triangulation during the initial phase of the MD simulation. The MD simulation and the LME regularization are however able to smooth out the local noise produced by such perturbations, as the simulation time progressively increases. 
}}

\begin{figure}[ht]
\unitlength1cm
\begin{picture}(14.5,10)
\if\Images y\put(0.75,0.0){\epsfig{figure=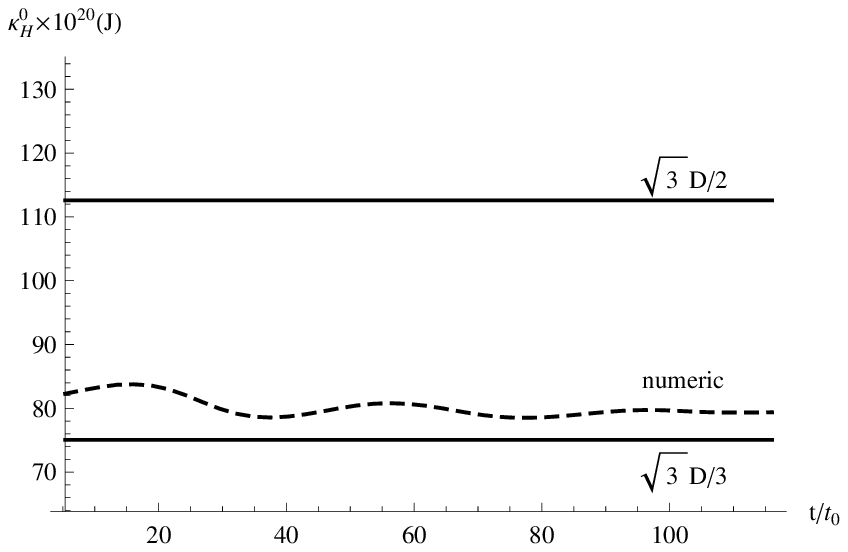,width=12cm}}\fi
\end{picture}
\caption{
{\cred{LME estimates of the zero-temperature bending rigidity  of the RBC membrane
model presented in Section \ref{RBC}, at different times of a MD simulation.}}}
\label{RBC_Bmod_history}
\end{figure}

{\cc{

\subsection{\label{bending_rigidity}Estimation of the bending rigidity of membrane networks from MD simulations}

Let us consider now a triangulated membrane network endowed with the following dihedral angle energy

\begin{align}\label{eq:E-bend}
\begin{split}
  E^{\rm dihedral} 
  &= D \sum_{\triangle,\triangle' \in {\cal C}_{\triangle} \atop \text{neighbors}} 
     \left(1 - \cos\widehat{\triangle\triangle'} \right) \ 
  = \  \frac{D}{2} \sum_{\triangle,\triangle' \in {\cal C}_{\triangle} \atop \text{neighbors}} |n_{\triangle} - n_{\triangle'}|^2
\end{split}
\end{align}
where ${\cal C}_{\triangle}$ denotes the set of all triangles forming the network;
$\widehat{\triangle \triangle'}$ is the dihedral angle between the triangles $\triangle$ and $\triangle'$; 
$n_{\triangle}$ is the unit normal to the triangle $\triangle$;
and the summation runs over all the pairs $\triangle,\triangle' \in {\cal C}_{\triangle}$  which share a common side. According to \citep{Seung:1988, Gompper:1996}, the limiting bending energy of such a network is a Helfrich-type bending energy \citep{Helfrich:1976} of the form

\beq
E^{{bend}} \ = \ \frac{\kappa_H}{2} \ \int_{S_N} ((2 H)^2  \ - \ 2 K) \ dS
\label{Ebend}
\eeq

\noindent where $\kappa_H$ is the bending rigidity, and it results $\kappa_H=\sqrt{3}D/3$ for a sphere, and $\kappa_H=\sqrt{3}D/2$ for a cylinder \citep{Gompper:1996} {\cb{(Note that in \citep{Seung:1988, Gompper:1996} the symbol $H$ is used to denote twice the mean curvature of the network)}}.

The LME prediction of the principal curvatures of the network allows us to estimate the limiting bending energy (\ref{Ebend}). 
As a matter of fact, the computing of the mean curvatures ${ H}_N=\{H_N^{a}, \ a=1,...,N\}$ of the current configuration leads us to approximate  (\ref{Ebend}) as follows

\beq
E_N^{{\rm bend}} \ = \ 
\frac{\kappa_H}{2} \ \left[
 \left( \sum_{a=1}^{N} \ (2 H_N^{a})^2 A_N^{a} \right) \ - \
8 \pi \right]
\label{Ebend_approx}
\eeq

\noindent where $A_N^{a}$ is the surface area of the $a$-th element of a dual tessellation of the network, which we assume is formed by polygons joining the midpoints of the network edges with the triangle barycenters (barycentric dual mesh). Equation (\ref{Ebend_approx}) can be employed to 
estimate
the bending modulus at zero temperature of the network, 
{\color{blue}{through
the Cauchy-Born contribution \citep{Ericksen:2008, Zhou:1996} to the isothermal bending rigidity, here denoted by $\kappa_H^0$.
The latter consists of }} the configurational average of the instantaneous bending rigidities of the fluctuating network. On 
matching $E_N^{{\rm bend}} $ to $E^{{\rm dihedral}} $ 
(energetic discrete-to-continuum approach), we identify the instantaneous bending rigidity with the ratio 
{\color{blue}{
$ E^{\rm dihedral} /  \left. \left( E_N^{\rm bend} \right) \right|_{\kappa_H = 1}$ }}, 
and compute $\kappa_H^0$ through
{\color{blue}{
\beq
\kappa_H^0 \ = \  \left\langle \frac{E^{{\rm dihedral}}}{ \left. \left( E_N^{\rm bend} \right) \right|_{\kappa_H = 1}}  \right\rangle
\label{KH0}
\eeq
}}
\noindent where $\langle \cdot \rangle$ denotes the configurational average symbol.

For the sake of example, let us reconsider now the MD model of the RBC membrane examined in the previous section. 
Fig. \ref{RBC_Bmod_history} shows the time history that we obtained for  the zero-temperature bending rigidity $\kappa_H^0$ of such a model,  on computing the rolling averages of the quantity ${2 \ E^{{\rm dihedral}}}/{E_N^{{\rm bend}}}$ during a MD simulation (cf. Sect. \ref{RBC}). It is seen from Fig. \ref{RBC_Bmod_history} that the value of $\kappa_H^0$  slightly fluctuates during the simulation, featuring oscillations with progressively smaller amplitude as the computational time $t$ increases. We estimated a limiting value of $\kappa_H^0$ approximatively equal to 79 J, for $t > 100 \ t_0$. Such a value is just slightly greater than that  predicted in \citep{Gompper:1996} for the sphere ($\sqrt{3}D/3=75.06$ J), which is not surprising, since the biconcave (average) shape of the RBC model under consideration has the same genus of the sphere.
}}

\bigskip


\section{Concluding remarks}
\label{conclusions}

{\cb{
We have presented and numerically tested a meshfree approach to the curvature and bending rigidity estimation of membrane networks, through a suitable extension of the LME method formulated in \citep{Arrojo:2006}.}}
The results of Sections \ref{sinus} and  \ref{sphere} {\cb{demonstrate the}} convergence properties of the LME curvature estimates, both for a rectangular geometry (fixed $\{ x_1,x_2,z \}$ axes), and for a spherical membrane. 
{\cred{
On the other hand, the results presented in Sections \ref{RBC} and \ref{bending_rigidity}, emphasize the ability of the LME approach in tracking the local curvatures and the bending rigidity of the RBC model presented in \citep{Marcelli:2005}. 
Concerning the parameter estimation, we have found that the limitations $\bar{\beta} \ge 100$ and $m \ge 10$ generally ensure stable curvature predictions.
In particular, for $m \ge 10$ we found that100 is approximately the smallest value of $\bar{\beta}$ that guarantees asymptotic convergence of LME curvature predictions for the examined examples.
}} {\cb{The value of $\bar{\beta}$}} {\cred{rules the degree of locality of LME approximations, which reduce to piecewise affine shape
functions supported by a Delaunay triangulation for $\bar{\beta} \rightarrow \infty$ \citep{Arrojo:2006, Cyron:2009}.
}}

In closing, we suggest some directions for relevant extensions of the present work.
{\cred{
We intend to  apply the LME regularization algorithm to predict the entire set of the isothermal membrane}} {\cb{and}} {\cred{bending rigidities of fluctuating biomembranes modeled through MD simulations, with special reference to the RBC membrane. We plan to combine the LME regularization of the RBC model proposed in 
\citep{Marcelli:2005},
with the elastic moduli estimation procedures given in
\citep{Seung:1988, Gompper:1996, Zhou:1996}.
}}
Another application of the LME procedure presented in this work is to mesure the curvature of lipid bilayers as modelled with MD simulations.  
Lipid bilayers are generally flat, however several different protein-driven processes will result in the membrane curvature that is required for various cell processes (i.e.\ fusion).  
One mechanism is caused by protein domains inserting shallowly into one of the lipid leaflets, which push the {\cb{neighboring}} lipid head groups away and therefore causing local spontaneous curvature.  We are currently conducting coarse-grain MD simulations that model the interactions between different varieties of these protein domains and lipid bilayers.  The LME procedure will allow us to quantify the amount of curvature that results from these interactions.

\section*{Acknowledgements}
F.\ F.\ wishes to acknowledge the great support received by Ada Amendola (Department of Civil Engineering, University of Salerno), Bo Li
(Graduate Aerospace Laboratories, California Institute of Technology),
and Franca Fraternali 
(Randall Division of Cell and Molecular Biophysics, King's College London)
during the course of the present work.

\bigskip

\bibliographystyle{apalike}  
\bibliography{thebibliography}	

\begin{thebibliography}{}

\bibitem[Arroyo and Ortiz, 2006]{Arrojo:2006}
Arroyo, M. and Ortiz, M. (2006).
\newblock Local maximum-entropy approximation schemes: a seamless bridge
  between finite elements and meshfree methods.
\newblock {\em Int. J. Numer. Meth. Eng.}, 65(13):2167--2202.

\bibitem[Cyron et~al., 2009]{Cyron:2009}
Cyron, C., Arrojo, M., and M., O. (2009).
\newblock Smooth, second-order, non-negative meshfree approximants selected by
  maximum entropy.
\newblock {\em Int. J. Num. Meth. Enging.}, 79:1605--1632.

\bibitem[Dao et~al., 2006]{Suresh:2006}
Dao, M., Li, J., and Suresh, S. (2006).
\newblock Molecularly based analysis of deformation of spectrin network and
  human erythrocyte.
\newblock {\em Mat. Sci. Engng.}, 26:1232--1244.

\bibitem[Deuling and Helfrich, 1976]{Helfrich:1976}
Deuling, H. and Helfrich, W. (1976).
\newblock Red blood cell shapes as explained on the basis of curvature
  elasticity.
\newblock {\em Biophys J.}, 16(8):861--868.

\bibitem[Du et~al., 2006]{Du:2006}
Du, Q., Liu, C., and Wang, X. (2006).
\newblock Simulating the deformation of vescicle membranes under elastic
  bending energy in three dimensions.
\newblock {\em J. Comput. Phys.}, 212:756--777.

\bibitem[Ericksen, 2008]{Ericksen:2008}
Ericksen, J. (2008).
\newblock On the cauchy-born rule.
\newblock {\em Math. Mech. Solids}, 13:199--220.

\bibitem[Gompper and Kroll, 1996]{Gompper:1996}
Gompper, G. and Kroll, D. (1996).
\newblock Random surface discretization and the renormalization of the bending
  rigidity.
\newblock {\em J. Phys. I France}, 6:1305--1320.

\bibitem[Hale et~al., 2009]{Hale:2009}
Hale, J., Marcelli, G., Parker, K., Winlowe, C., and Petrov, G. (2009).
\newblock Red blood cell thermal fluctuations: comparison between experiment
  and molecular dynamics simulations.
\newblock {\em Soft Matter}, 5:3603--3606.

\bibitem[Jaines, 1957]{Jaines:1957}
Jaines, E. (1957).
\newblock Information theory and statistical mechanics.
\newblock {\em Phis. Rev.}, 106:620--630.

\bibitem[Li et~al., 2010]{Li:2010}
Li, B., Habbal, F., and Ortiz, M. (2010).
\newblock Optimal transportation meshfree approximation schemes for fluid and
  plastic flows.
\newblock {\em Int. J. Num. Meth. Enging.}

\bibitem[Marcelli et~al., 2005]{Marcelli:2005}
Marcelli, G., Parker, H., and Winlove, P. (2005).
\newblock Thermal fluctuations of red blood cell membrane via a constant-area
  particle-dynamics model.
\newblock {\em Biophys. J.}, 89:2473--2480.

\bibitem[Martens and McMahon, 2008]{Martensetal.:2008}
Martens, S. and McMahon, H. (2008).
\newblock Mechanisms of membrane fusion: disparate players and common
  principles.
\newblock {\em Nature Rev.}, 9:543--556.

\bibitem[McMahon and Gallop, 2005]{McMahonetal.:2005}
McMahon, H. and Gallop, J. (2005).
\newblock Membrane curvature and mechanisms of dynamic cell membrane
  remodelling.
\newblock {\em Nature}, 438:590--596.

\bibitem[Naghdi, 1972]{Naghdi72}
Naghdi, P. (1972).
\newblock The theory of shells and plates.
\newblock In Trusdell, C., editor, {\em S. Fl{\"u}gge's Handbuch der Physik},
  volume VIa/2, pages 425--640. Springer Verlag.

\bibitem[Rajan, 1994]{Rajan:1994}
Rajan, V. (1994).
\newblock Optimality of the delunay triangulation in $r^d$.
\newblock {\em Discrete Comput. Geom.}, 12(2):189--202.

\bibitem[Risselada and Marrink, 2009]{RisseladaetMarrink:2009}
Risselada, H.~J. and Marrink, S. (2009).
\newblock Curvature effects on lipid packing and dynamics in liposomes revealed
  by coarse grained molecular dynamics simulations.
\newblock {\em Phys. Chem.}, 11:2056--2067.

\bibitem[Seung and Nelson, 1988]{Seung:1988}
Seung, H. and Nelson, D. (1988).
\newblock Defects in flexible membranes with crystalline order.
\newblock {\em Phys. Rev. A}, 38:1005--1018.

\bibitem[Smith and Forester, 1999]{Smith:1999}
Smith, W. and Forester, T. (1999).
\newblock The dl\_poly\_2 molecular simulation package.
\newblock {\em http://www.cse.clrc.ac.uk/msi/software/DL\_POLY}.

\bibitem[Stoker, 1969]{Stoker:1969}
Stoker, J.~J. (1969).
\newblock {\em Differential Geometry}.
\newblock Wiley, New York.

\bibitem[Taubin, 1995]{Taubin:1995}
Taubin, G. (1995).
\newblock Estimating the tensor of curvature of a surface from a polyhedral
  approximation.
\newblock In {\em Proc. 5th Intl. Conf. on Computer Vision (ICCV95)}, pages
  902--907.

\bibitem[Zhou and Jo{\'o}s, 1996]{Zhou:1996}
Zhou, Z. and Jo{\'o}s, B. (1996).
\newblock Stability criteria for homogeneously stressed materials and the
  calculation of elastic constants.
\newblock {\em Phys. Rev. B}, 54(6):3841--3850.

\end{thebibliography}

\end{document}